\documentclass{aa}
\usepackage{graphics}
\usepackage{natbib}
\usepackage{graphicx}
\usepackage{amssymb}
\begin{document}

\title{X-ray properties of active M dwarfs as observed by XMM-Newton}

\author{J. Robrade and J.H.M.M. Schmitt}
\institute{
Hamburger Sternwarte, Universit\" at Hamburg, Gojenbergsweg 112,
D-21029 Hamburg, Germany}

\authorrunning{Robrade \& Schmitt}
\titlerunning{M dwarfs in X-rays}
\offprints{J. Robrade}
\mail{jrobrade@hs.uni-hamburg.de}
\date{Received 3 September 2004 / Accepted 8 February 2005}

\abstract{
We present a comparative study of X-ray emission from a sample of active M~dwarfs
with spectral types \hbox{M3.5\,-\,M4.5} using XMM-Newton observations
of two single stars, AD~Leonis and EV~Lacertae, and two unresolved
binary systems, AT~Microscopii and EQ~Pegasi. 
The light curves reveal frequent flaring during all four observations.
We perform a uniform spectral analysis and determine plasma temperatures, abundances and
emission measures in different states of activity.
Applying multi-temperature models with variable abundances separately to
data obtained with the EPIC and RGS detectors we are able to
investigate the consistency of the results obtained by 
the different instruments onboard XMM-Newton. 
We find that the X-ray properties of the sample M~dwarfs are very similar, 
with the coronal abundances of all sample stars 
following a trend of increasing abundance with increasing first ionization potential,
the inverse FIP effect.
The overall metallicities are below solar photospheric ones but appear
consistent with the measured photospheric abundances of M~dwarfs like these.
A significant increase in the prominence of the hotter plasma components 
is observed during flares while the cool plasma component is only marginally affected by flaring,
pointing to different coronal structures.
AT~Mic, probably a young pre-main-sequence system, has the highest X-ray luminosity 
and exhibits also the hottest corona. 
While results of EQ~Peg and EV~Lac are presented here for the first time,
AT~Mic and AD~Leo have been investigated before with different analysis approaches, allowing
a comparison of the results.
}
\maketitle

\section{Introduction}
\label{intro}
Observations with ROSAT
have shown the formation of X-ray emitting coronae around late-type cool dwarf 
stars with outer convection zones to be universal \citep[][]{schmitt95, schmitt04}.
Coronal structures on the Sun are dominated by their magnetic properties
with closed structures, the coronal loops or loop arcades, containing predominantly the X-ray emitting,
confined hot high-density plasma and being the location of flare events.
In the Sun the interaction between the
radiative and the outer convective zones powers a dynamo, leading to coronal activity.
The ubiquitous occurrence of X-ray emission among cool stars and their dependence on 
rotation strongly suggests also a magnetic character of their activity.
Observed activity levels of cool stars, as measured by the ratio of bolometric and X-ray luminosities,
span a range of values from $\sim10^{-8}$ to $\sim10^{-3}$ that are correlated with dynamo efficiency.
The efficiency of a solar-like magnetic dynamo is characterised by the inverse of
the so-called 'Rossby-Number', defined as the ratio between rotational period and convective turnover time.
This activity-rotation relation saturates at $L_{\rm X}$/$L_{\rm bol}$\,$\sim10^{-3}$, a level~$>$\,1000 times
higher than that of the Sun.
The similarity of stellar and solar X-ray emission and flares suggests similar basic physical mechanisms, with
more active stars showing more frequent and larger flares.
Flaring has been proposed as a possible
major coronal heating mechanism for the Sun via nanoflares \citep{parker} as well as for 
other cool stars including M~dwarfs.
For recent reviews on this topic, see \cite{flares1} or \cite{flares2}.

Active M~dwarfs, belonging to the so-called flare stars, 
turned out to be particularly strong coronal X-ray sources. 
Frequent flaring on these objects was first discovered in the optical regime, 
later in EUV and X-rays. These late-type dwarf stars are generally
more active than solar-like stars since they have larger spin-down times. 
They also typically show higher coronal temperatures.
M~dwarfs of spectral types later than approximately \hbox{M3\,-\,M5} are expected to be fully convective
and therefore lack an interface between radiative and convective zones.
The observed high levels of activity then require the onset of alternative dynamo mechanisms different to that 
of the Sun.

The stars chosen here for analysis are known high X-ray emitters
with spectral types in the range \hbox{M3.5\,-\,M4.5} 
and are located in the immediate solar vicinity (distance 5\,-\,10 pc). 
EQ~Peg (\object{Gl~896A}, \object{Gl~896B}) is a visual binary consisting 
of two M~dwarfs of spectral type M3.5 and M4.5, both of which are flare stars, 
as well as AT~Mic (\object{Gl~799A}, \object{Gl~799B}), an active binary system consisting of a M4.5 and a M4 star. 
Both binaries are separated by roughly 30\,AU, making interaction between the components quite unlikely.
On the other hand, AD~Leo (\object{Gl~388}) is a single M3.5 star similar to EV~Lac (\object{Gl~873}).
Our sample M~dwarfs are also all fast rotators ($ v \sin i\,>\,$5\,km/s) \citep{rot1,rot2} 
in the saturated regime, i.e $L_{\rm X}$/$L_{\rm bol}$\,$\sim10^{-3}$.
Stellar parameters important for this work are summarized in Table \ref{stars}.
Note that the spectral classification in the literature is not unique for all M~dwarfs 
in our sample and may differ by one spectral subclass; 
the values listed here are taken from the SIMBAD database.

\begin{table}[!ht]
\setlength\tabcolsep{4pt}
\caption{\label{stars}Stellar parameters; vsini from \cite{rot1} except AT~Mic \citep{rot2} and EQ~Peg \citep{stau86}, 
J from 2MASS point source catalog, $L_{bol}$ calculated from J using values of \cite{jcorr}, $L_{bol}$/$L_{X}$ 
with $L_{X}$ (this work, PN).}
{\scriptsize
\begin{tabular}{lccccc}\hline\hline
Star & d(pc) & vsini(km/s) &J(mag) & log$L_{bol}$ & log$L_{bol}$/$L_{X}$ \\\hline
AT Mic A+B & 10.2 & 11.7(A) & 5.81 & 32.46 &  -3.00 \\
EQ Peg A+B & 6.2 & 12/18(A/B) & 5.79 & 32.04 & -3.22\\
AD Leo & 4.7 & 6.2 & 5.45 & 31.94 & -3.21\\
EV Lac & 5.0 &6.9 &6.11 &31.73 &-2.99\\\hline
\end{tabular}
}
\end{table}

While X-ray emission from AD~Leo and AT~Mic was already detected by HEAO\,1 \citep{heao1} and
EQ~Peg is contained in the {\it Einstein} Stellar Survey \citep{vai81},
EV~Lac was first detected by EXOSAT \citep{schmitt88}.
All stars in our sample were also detected in the ROSAT all-sky survey \citep{huensch99} 
and were observed by various other X-ray missions.
A comparative analysis of X-ray spectra from AD~Leo and EV~Lac obtained with BeppoSAX and 
ROSAT PSPC was performed by \cite{sax}. \cite{favata00a} presented 
a study of an extreme X-ray flare on EV~Lac observed with ASCA.
\cite{favata00b} studied AD~Leo data taken by {\it Einstein}, ROSAT and ASCA and found the
coronal structures to be compact and the quiescent coronal luminosity remarkably constant over decades.

Parts of the XMM-Newton data of AT~Mic were previously analysed with 
special focus on elemental abundances and
emission measures during flaring and quiescent state \citep{atmic}.  
While the coronae of low activity stars like the Sun \citep{laming} show an enhancement of
low-FIP elements, the so-called FIP (First Ionization Potential) effect, \cite{atmic} discovered
in the corona of the highly active star AT~Mic an inverse FIP effect, which appears to flatten during the flare state.
A comparative study of flares on AD~Leo 
observed with XMM-Newton and {\it Chandra} was performed by \cite{adleo} and again an inverse FIP effect
was found. An analysis of spatially resolved X-ray emission of the EQ~Peg system was presented by \cite{eqpeg}.
A comparative study on coronal density diagnostics with He-like triplets 
that includes work on our sample stars
can be found in \cite{helike} and with special focus on the X-ray sizes of coronae in \cite{helike1}. 

The above listed M~dwarfs were observed with XMM-Newton and the goal of this paper is to present 
a comparative analysis of these X-ray observations and to determine those properties that are typical for
active M~dwarfs as a class and those attributed to individual objects.
In Sect.\,\ref{anal} we describe the observations and the methods used for data analysis. 
In Sect.\,\ref{results} we present the results subdivided into different physical topics which is
followed by a summary and our conclusions in Sect.\,\ref{summ}.

\section{Observation and data analysis}
\label{anal}

The four targets, AD~Leo, AT~Mic, EV~Lac and EQ~Peg, were observed  
with XMM-Newton using varying detector setups and 
exposure times in the range of 15\,-\,35\,ksec. No attempts were made to
separate the binaries in this work,
therefore data from AT~Mic and EQ~Peg is a superposition of the respective system components.
A detailed description of the observations is provided in Table~\ref{obs}. 
Useful data were collected in all X-ray detectors onboard XMM-Newton, respectively
the EPIC (European Photon Imaging Camera) and RGS (Reflection Grating Spectrometer) detectors
which are operated simultaneously.
The EPIC instrument consists of three CCD cameras with two different CCD designs, 
respective two MOS and one PN, providing imaging and spectroscopy in the energy range from 0.15 to 15\,keV
with good angular and moderate spectral resolution. 
In this work EPIC data is used only when the detectors observed in the imaging mode
and therefore data from both MOS detectors are presented only for EQ~Peg .
The RGS is a grating spectrometer with high spectral resolution consisting of two identical 
spectrometers. The RGS operates in the energy range from 0.35 to 2.5\,keV (5\,-\,35\,\AA) with a 
spectral resolution of $\sim$\,4.0\,eV (0.06\,\AA, FWHM) at 1.0\,keV in first order.
A detailed description of the XMM-Newton instruments is given by \cite{xmm}.

\begin{table}[!ht]
\caption{\label{obs}Observation log}
{\scriptsize
\begin{tabular}{llcc}\hline\hline
 Target & Instrument (Mode) & Dur. (s) & Obs. Time\\\hline
 EQ Peg & MOS (FF,thick F.) & 14600 & 2000-07-09T11:39-15:43 \\
   "    & PN  (FF,thick F.) & 12410 & 2000-07-09T12:20-15:47 \\
   "    & RGS (HER/SES)     & 15610 & 2000-07-09T11:31-15:51 \\\hline
 AT Mic & MOS1 (SW,med. F.) & 25400 & 2000-10-16T00:28-07:32 \\
   "    & PN  (SW,med. F.) & 25100 & 2000-10-16T00:42-07:40 \\
   "    & RGS (HER/SES)    & 28310 & 2000-10-16T00:20-08:11 \\\hline
AD Leo & MOS2 (LW,thick F.) & 35800 & 2001-05-14T20:55-06:52 \\
   "    & PN  (SW,med. F.) & 35000 & 2001-05-14T21:11-06:54 \\
   "    & RGS  (HER/SES)    & 36400 & 2001-05-14T20:48-06:55\\\hline
EV Lac & MOS2 (SW,med. F.) & 32260 & 2001-06-03T07:55-16:53 \\
   "    & PN  (SW,med. F.) & 31470 & 2001-06-03T08:11-16:55 \\
   "    & RGS  (HER/SES)   & 32860 & 2001-06-03T07:49-16:56\\\hline
\end{tabular}
}
\end{table}

The data were reduced with the standard XMM-Newton Science Analysis System (SAS)
software, version 5.4.1 with calibration files updated in January 2004. 
Light curves and spectra were produced with standard
SAS tools and standard selection criteria were applied for filtering the data \citep[see][]{sas}.
Spectral analysis was done with XSPEC V11.3 \citep{xspec}.

Data from the detectors of different type were fitted separately to allow comparison of the 
results and to check for possible cross-calibration problems. 
Spectral analysis of EPIC data is performed in the energy band between 0.2\,-\,12.0\,keV.
RGS first order spectra in the full energy range, 
i.e. 0.35\,-\,2.5\,keV (5\,-\,35\,\AA) are used.
Data of the same detector type, e.g. RGS1 and RGS2, and in the EQ~Peg case MOS1 and MOS2,
were analyzed simultaneously but not co-added. 
For spectral analysis 
the data was cleaned of proton flares by removing the affected time periods, and
in the EQ~Peg case the data was additionally cleaned for pile-up by excluding the inner part of the source emission.
The applied selection criteria were adjusted according to observation mode.
The background was taken from source-free regions on the detectors.

We apply the same type of spectral model to all data to perform a self-consistent analysis.
Therefore we use the global fitting approach, where the whole spectrum is fitted instead 
of a line-based method which would
be applicable only to RGS data. \cite{aud04} applied both methods to data of 
YY Mensae taken by {\it Chandra} and XMM-Newton and
found consistent results. These authors fitted data from RGS and MOS in
selected energy ranges simultaneously, similar to our joint fits, which are briefly discussed in Sect.\ref{specfits}. 
Larger deviations were found when using data of different detectors, a problem also
affecting our analysis. However, \cite{ab1} analysed {\it Chandra} and XMM-Newton data of AB~Dor, 
a highly active near-main-sequence K star,
and found the line-based method to be better suited to fit their high resolution spectra, leading to slightly
higher abundances and lower emission measures to be present in the derived models compared to the global
fitting approach. \cite{ab2} analysed EPIC and RGS spectra from different observations 
of AB~Dor using the global approach and 
found different absolute abundance values but a pattern 
(increasing depletion of abundance of low FIP elements up to iron followed
by the inverse pattern for elements with higher FIP) overall consistent with the results derived by \cite{ab1},
demonstrating the robustness of abundance ratios in general.

For the analysis of the X-ray spectra we specifically use multi-temperature models with variable but tied abundances,
i.e. the same abundance pattern in all temperature components.
Such models assume the emission spectrum of a collisionally-ionized diffuse 
optically-thin gas as calculated with the APEC code \citep[see][]{apec}. 
Abundances are calculated relative to solar photospheric values as given by 
\cite{and89}. For iron and oxygen we use the updated values of \cite{grev98}.
Comparing stellar coronal with solar photospheric abundances is of course crude, but precise photospheric
abundances are unknown for the observed targets; 
a detailed discussion of the abundances is given in Sect.\,\ref{abusec}.
Due to the proximity of the stars, absorption in the interstellar medium is negligible at 
the wavelengths of interest and was not applied in our modelling.
Our fit procedure is based on $\chi^2$ minimization, therefore spectra 
are rebinned to satisfy the statistical demand of a minimum value of counts per spectral 
bin; 25 for data with an average count rate above 5\,cts\,s$^{-1}$
and 15 for data with an average count rate below 5\,cts\,s$^{-1}$ and for all data used in time resolved analysis.
Bad channels are always excluded from the fit; these include channels at high energy in the EPIC spectra that
contain insufficient source counts (see Fig.\,\ref{mdspecpn}).
The first model uses three temperature components (\hbox{3\,-T} model) with free temperatures and emission measures.
Models with additional temperature components were checked, but did not improve the fit results significantly.
In order to approximate physically more realistic continuous temperature distributions of the plasma, 
we use a \hbox{6\,-T} model on a logarithmic, almost equidistant grid
(0.2,\,0.3,\,0.6,\,1.2,\,2.4,\,4.8\,keV), which covers roughly the temperature range from 2 to 60\,MK 
and samples those spectral regions where the XMM-Newton detectors are most sensitive.

Consider next the different spectral resolution, $\Delta E/E$, and energy range of the EPIC and RGS instruments.
Especially at low energies, $\Delta E/E$ is quite large for the EPIC detectors,
while in the RGS most of the strongest emission lines are separated in this energy range, thus
permitting a more accurate determination of plasma parameters.
Some elements, in particular C and N, show prominent features only in the RGS.  
Therefore, to fit the EPIC data the abundances of these elements were set to the RGS value 
in order to avoid unphysical solutions.
For elements such as Al, Ca and Ni, whose abundances in the solar photosphere are
an order of magnitude lower than that of Fe and which show 
only very weak features in the X-ray spectrum, we adopted a coronal abundance 
of 0.4 times the solar photospheric abundance (a typical value for Fe in this sample).
The EPIC detectors are able to measure higher energy X-rays, 
important for a reliable determination of the properties of the hotter plasma components, with
the MOS detectors providing a slightly better spectral resolution and the PN detector
providing greater sensitivity.

An important feature of the considered spectral models is
the interdependency between the emission measure (EM=$\int n_{e}n_{H}dV$) of temperature components 
on the one hand and elemental abundances with temperature sensitive emission 
lines in the considered temperature range on the other hand.
This is especially true for EPIC data with their
moderate spectral resolution.
Another general source of uncertainties is the incomplete knowledge of 
atomic data used to calculate the spectral models, see e.g. \cite{porquet} and references therein.

\section{Results}
\label{results}
\subsection{Investigation of light curves}
\label{lcana}

\begin{figure}[!ht]
\includegraphics[width=88mm,height=55mm]{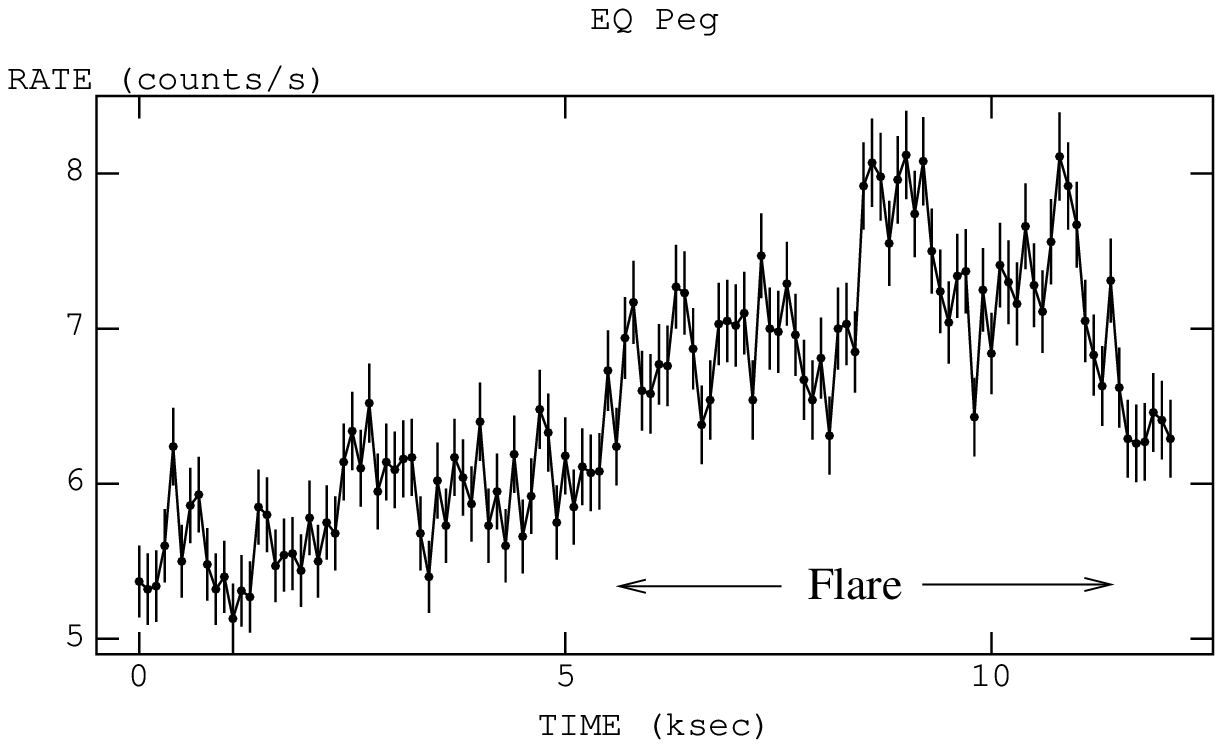}
\vspace*{-5mm}

\includegraphics[width=88mm,height=55mm]{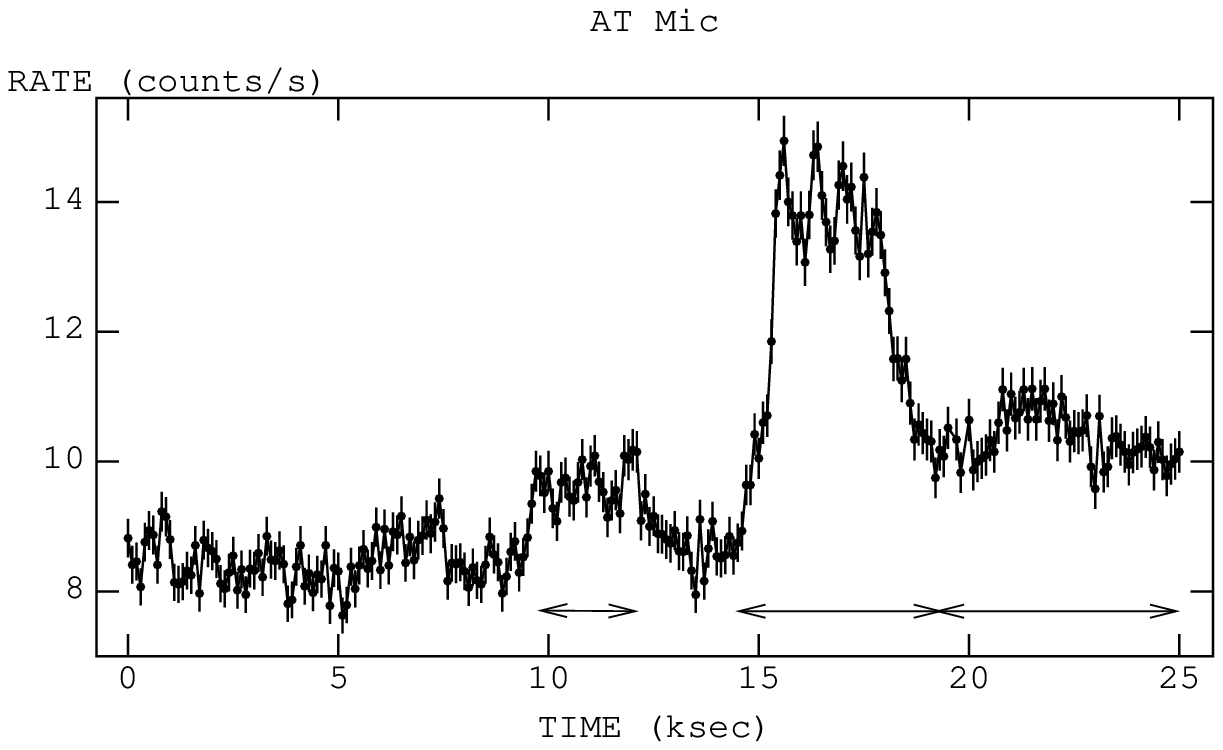}
\vspace*{-5mm}

\includegraphics[width=88mm,height=55mm]{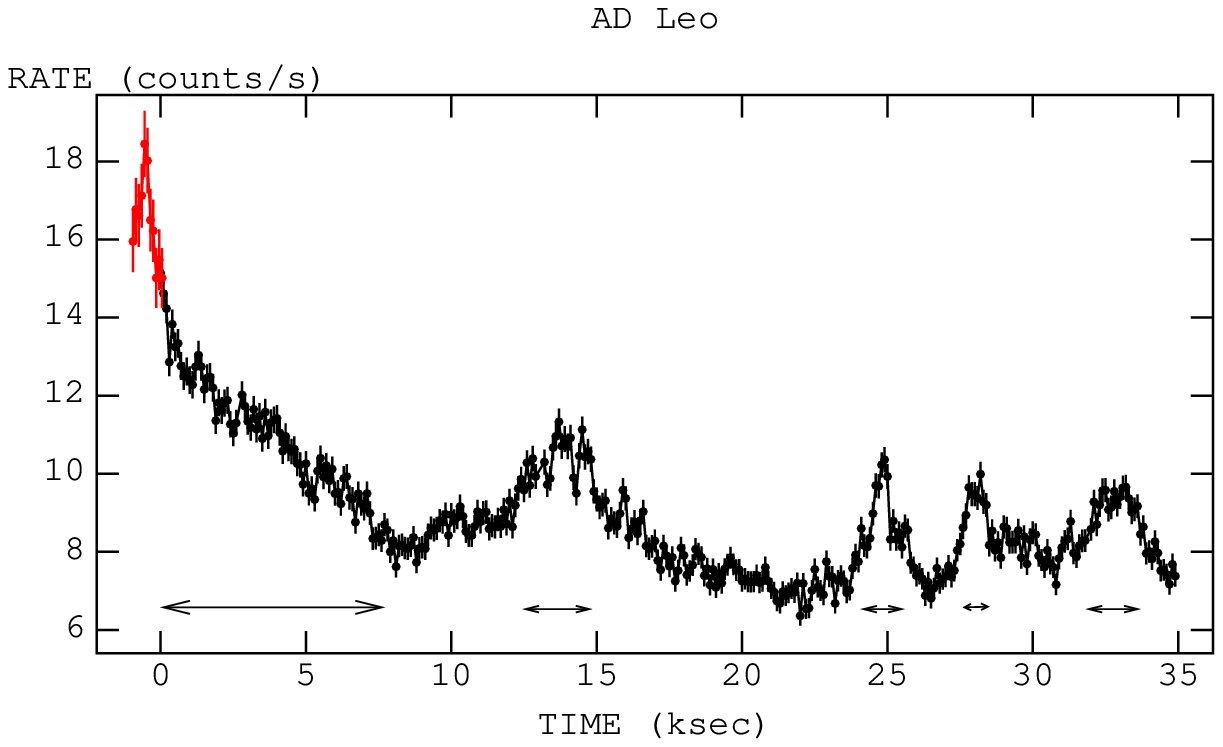}
\vspace*{-5mm}

\includegraphics[width=88mm,height=55mm]{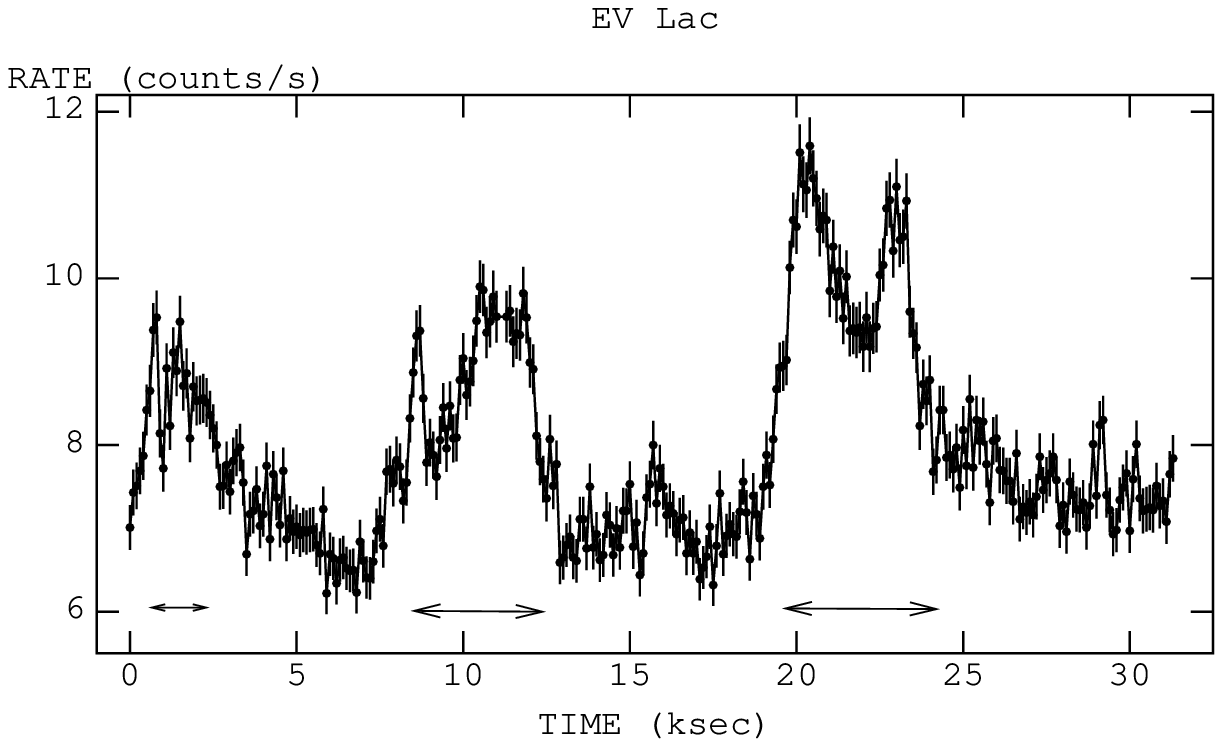}
\caption{\label{lc1}Light curves of the binaries EQ~Peg and AT~Mic and the single stars
 AD Leo and EV Lac; PN (+MOS2/red) data, 100\,s binning. Frequent flaring is observed on various timescales;
selected flaring periods are indicated by horizontal arrows.}
\end{figure}

The PN light curves of the four observations are shown in Fig.\,\ref{lc1};
these light curves were extracted from a circle with 50\,\arcsec\,radius around each source
and the temporal binning is 100\,s for all sources. 
Light curves were cleaned (AD~Leo, AT~Mic, EV~Lac) 
for obvious data dropouts and high proton flare background was subtracted from the light curve (EV~Lac). 
As the PN exposure of AD~Leo began during a flare, we included scaled data from the MOS2 exposure,
which started $\sim$ 1\,ksec earlier, to show the peak of the flare.

The light curves of the individual targets show obvious variability in all sources. 
The largest flares that occurred in these observations involve X-ray flux increases
of factors 2--3, which is considered moderate in view of flux increases 
of factors 10--300 that have been observed in flares on 
EQ~Peg \citep{kat02} and EV~Lac \citep{favata00a}. For an analysis of a larger
flare on the M~dwarf Proxima Centauri with XMM-Newton see \cite{proxcen}. 
While our observations indicate that flaring with flux increases of factors 2--3
appears regularly on timescales of a few hours, these larger events are much rarer. 

Weaker flaring (flux changes of around
ten percent or lower) on timescales of a few minutes is visible throughout all four observations. 
In this phase the stars are in a state of low but permanent activity. 
This continual weak flaring makes it, in addition to the larger flares, 
difficult to determine a minimum or baseline flux level.
This flux or activity level is denoted here as the quasi-quiescent level. 
It has been suggested that this emission can be interpreted as a superposition of a 
large number of even smaller flares with a power law distribution
\citep{flares3,flares1,flares2}.

The flare light curves exhibit a variety of shapes and often show substructure. The observed
rise times are only somewhat shorter than the decay times for the larger impulsive flares during 
the AT~Mic and the EV~Lac observation, pointing to relatively compact flaring structures.
The different types of flares as observed on the Sun are 
described in e.g. \cite{feldman} and references therein. 
The rise times from quasi-quiescent to flare maximum are typically 
in the range of 1\,ksec for the largest flares with
the rise steepening towards maximum.
The binaries are not resolved in these light curves, but all components are expected to be flare stars.
For EQ~Peg it was shown by \cite{eqpeg} that both stars contribute to the flaring, 
with EQ~Peg~A being brighter than B by a factor of 3\,-\,4.
The separation of AT~Mic is too small to resolve the components. 

Flaring involves significant amounts of energy release and matter transfer into the corona and may therefore involve
changes in the physical properties of the coronal plasma, e.g. temperatures, emission measures, densities 
and abundances, a well known phenomenon from previous analyses of stellar flares. 
Therefore we introduce 'flaring' and 'quasi-quiescent' phases for each star;
an interval is selected as flaring if the count rate of the time bins exceed the median count rate and the 
peak count rate within the interval exceeds a 'minimum flare value', 
chosen to separate clear flare events and flickering. 
The selection of two phases of activity is of course somewhat arbitrary,
and observations with stronger flares may be
separated into three or more phases of in- or decreasing activity. 
With our data this was only done for the AT~Mic observation, which is
separated in three states of activity, the quasi-quiescent phase (QQ), 
the medium activity phase and the outstanding large flare.
The count rates, selection criteria and selected flaring periods
are summarized in Table~\ref{selflare}.

\begin{table}[!ht]
\setlength\tabcolsep{6pt}
\caption{\label{selflare}Selection of flaring data (PN).}
{\scriptsize
\begin{tabular}{lcccc}\hline\hline
& EQ Peg & AT Mic & AD Leo & EV Lac \\\hline
Median rate & 6.42 & 9.40 & 8.60 & 7.58 \\
Min. flare rate & 7.0 & 12.0 (high) & 9.5 & 8.5\\
& & 10.0 (med) & & \\
Flare intervals & 5.5-11.6 & 14.7-19.1 (h) & 0.0-7.7 & 0.5-2.3\\
(ksec after & & 9.7-12.1 (m) & 12.2-14.9 & 8.4-12.2\\
start of & & 19.2-25.0 (m) &  24.1-25.7 & 19.4-24.0 \\
observation)& & &  27.4-28.4 & \\
& & &  32.0-33.8  & \\\hline
\end{tabular}
}
\end{table}

\subsection{Spectral analysis}
\label{specana}
\subsubsection{RGS and EPIC spectra}

\begin{figure}[!ht]
\includegraphics[width=90mm,height=80mm]{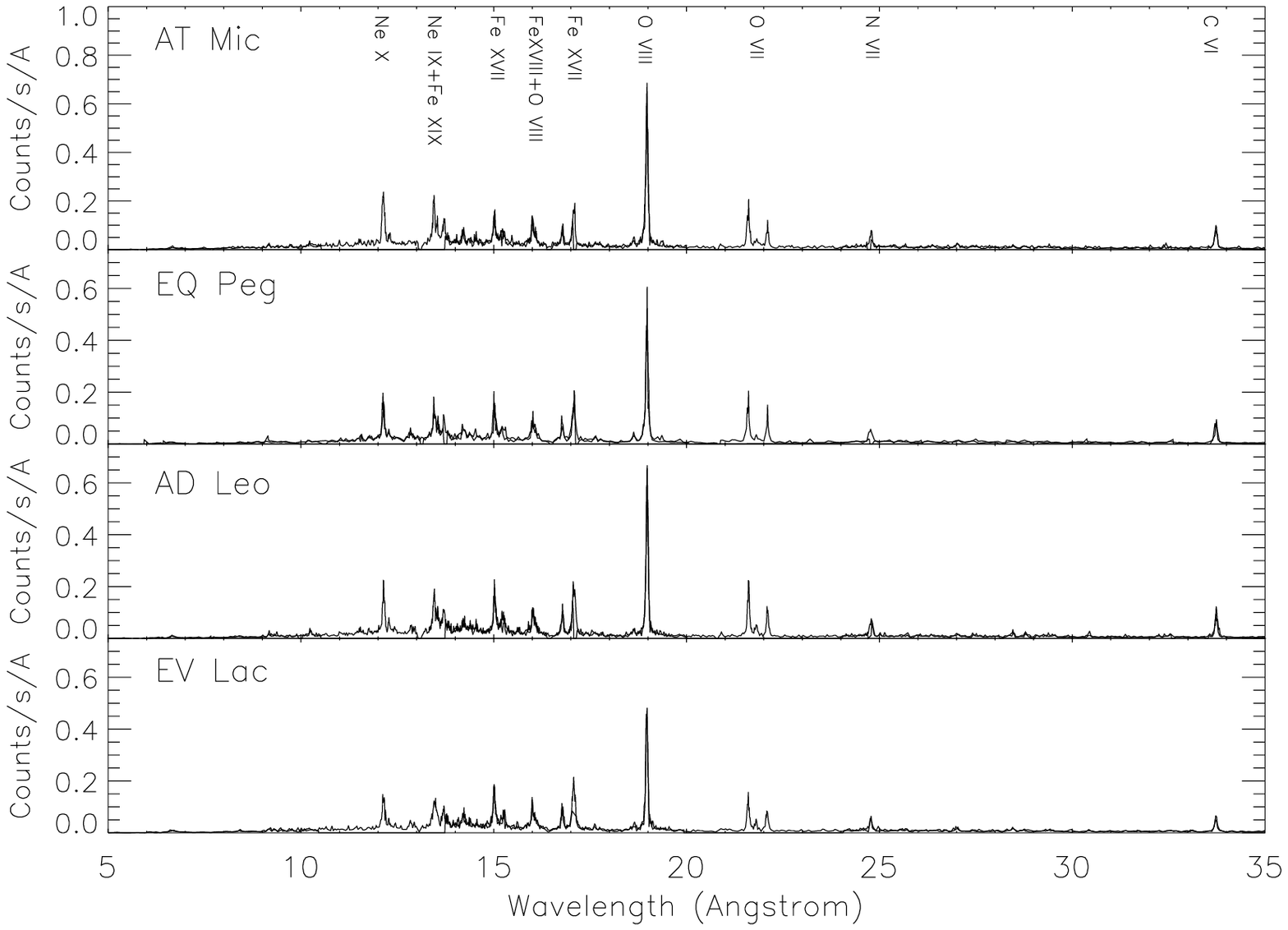}
\caption{\label{rgsspec}RGS spectra of the sample stars, the most 
prominent emission features are labelled.}
\end{figure}

The high resolution RGS spectra of our sources in the wavelength range 
5\,-\,35\,\AA \, are shown in Fig.\,\ref{rgsspec}.
In this energy range
the X-ray spectra of our M~dwarfs are remarkably similar, all being strongly dominated by emission lines.
The positions of the most prominent emission features are labelled; i.e., the H-like lines of \ion{C}{vi}, \ion{N}{vii}, 
\ion{O}{viii} and \ion{Ne}{x}, the strong He-like triplets of \ion{O}{vii} and \ion{Ne}{ix} as well as
a number of \ion{Fe}{xvii}\,-\,\ion{Fe}{xix} lines.
The peak formation temperatures of these lines 
cover temperatures from $\sim$\,1.5\,MK ( \ion{C}{vi}) over $\sim$\,3\,MK (\ion{O}{viii}) and $\sim$\,6\,MK (\ion{Ne}{x})
up to $\sim$\,8\,MK (\ion{Fe}{xix}) but 
plasma with temperatures up to 30\,MK can contribute to these lines.

\begin{figure}[!ht]
\includegraphics[width=56mm,angle=-90]{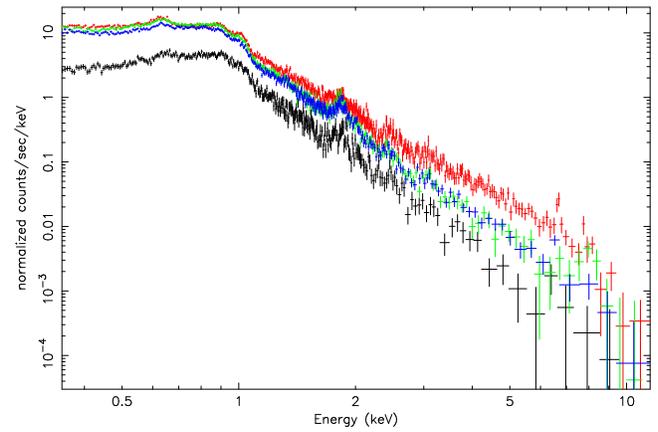}
\caption{\label{mdspecpn}PN spectra of the analysed M~dwarfs; top to bottom: AT~Mic (red), AD~Leo (green), 
EV~Lac (blue), EQ~Peg (black).[{\it Colour figure in electronic version.}]}
\end{figure}

The EPIC PN data, with moderate spectral resolution, are displayed
in Fig.\,\ref{mdspecpn} for our target stars; note that the 
EQ~Peg EPIC spectra were extracted from an annulus because of pile up. 
In addition to the features mentioned in the RGS section
the continuum at high energies and lines due to
highly ionized iron (up to \ion{Fe}{xxv} at 6.7\,keV) and H-like/He-like
magnesium, silicon and sulfur are detectable.
The observed features cover roughly the temperature range from 2 to 70\,MK 
in terms of peak formation temperatures.
Inspection of the PN spectra of the sample M~dwarfs
shows differences in the high energy slopes, the obviously shallower slope
of the AT~Mic spectrum indicates the higher coronal temperatures of this source.

\subsubsection{Spectral fits and coronal temperatures}
\label{specfits}

\begin{table*}[!ht]
\setlength\tabcolsep{5pt}
\caption{\label{res}Results on temperatures (keV), emission measures ($10^{51}\,$cm$^{-3}$) 
and abundances relative to solar photospheric values, errors are 90\,\% conf. range. (d.o.f. given in brackets).
$L_{\rm X}$ ($10^{28}\,$erg\,s$^{-1}$) indicates the 0.2\,-10.0\,keV luminosity. 
The 6-T model grid is 0.2,\,0.3,\,0.6,\,1.2,\,2.4,\,4.8\,keV which
corresponds to 2.3,\,3.5,\,7.0,\,14.0,\,28.0,\,56.0\,MK.}
{\scriptsize
\begin{tabular}{l|ccccc|cccccc}\hline\hline
Par. & RGS & MOS & PN & PN (low) & PN (high) & RGS & MOS & PN & PN (low) & PN (high)\\\hline
 & \multicolumn{5}{c}{EQ Peg  -  3-T model}& \multicolumn{5}{c}{EQ Peg  -  6-T model}\\\hline
C  & 0.71$^{+ 0.17}_{- 0.14}$ & -- & --  & -- & --  & 0.70$^{+ 0.18}_{- 0.13}$ & -- & --  & -- & --\\
N  & 0.69$^{+ 0.19}_{- 0.16}$ & -- & --  & -- & --  & 0.68$^{+ 0.22}_{- 0.15}$ & -- & --  & -- & --\\
O  & 0.54$^{+ 0.09}_{- 0.08}$ & 0.37$^{+ 0.05}_{- 0.04}$ & 0.58$^{+ 0.09}_{- 0.08}$  & 0.70$^{+ 0.18}_{- 0.14}$& 0.48$^{+ 0.10}_{- 0.07}$  & 0.59$^{+ 0.11}_{- 0.06}$ & 0.37$^{+ 0.06}_{- 0.05}$ & 0.58$^{+ 0.09}_{- 0.08}$  & 0.73$^{+ 0.18}_{- 0.15}$& 0.48$^{+ 0.08}_{- 0.07}$\\
Ne  & 0.96$^{+ 0.16}_{- 0.13}$ & 0.73$^{+ 0.14}_{- 0.12}$ & 0.93$^{+ 0.19}_{- 0.18}$  & 0.92$^{+ 0.32}_{- 0.37}$& 1.01$^{+ 0.27}_{- 0.22}$  & 0.96$^{+ 0.15}_{- 0.13}$ & 0.69$^{+ 0.09}_{- 0.109}$ & 1.00$^{+ 0.16}_{- 0.13}$  & 0.96$^{+ 0.26}_{- 0.21}$& 1.11$^{+ 0.20}_{- 0.17}$\\
Mg  & 0.30$^{+ 0.14}_{- 0.14}$ & 0.32$^{+ 0.08}_{- 0.07}$ & 0.25$^{+ 0.11}_{- 0.09}$  & 0.22$^{+ 0.19}_{- 0.15}$& 0.27$^{+ 0.15}_{- 0.11}$  & 0.27$^{+ 0.15}_{- 0.13}$ & 0.33$^{+ 0.08}_{- 0.08}$ & 0.28$^{+ 0.11}_{- 0.09}$  & 0.22$^{+ 0.17}_{- 0.14}$& 0.33$^{+ 0.14}_{- 0.12}$\\
Si  & 0.58$^{+ 0.31}_{- 0.29}$ & 0.58$^{+ 0.10}_{- 0.09}$ & 0.40$^{+ 0.12}_{- 0.10}$  & 0.55$^{+ 0.23}_{- 0.18}$& 0.29$^{+ 0.15}_{- 0.11}$  & 0.56$^{+ 0.34}_{- 0.28}$ & 0.60$^{+ 0.10}_{- 0.11}$ & 0.45$^{+ 0.13}_{- 0.11}$  & 0.58$^{+ 0.23}_{- 0.19}$& 0.37$^{+ 0.15}_{- 0.13}$\\
S  & 0.27$^{+ 0.18}_{- 0.15}$ & 0.68$^{+ 0.17}_{- 0.16}$ & 0.27$^{+ 0.21}_{- 0.20}$  & 0.63$^{+ 0.40}_{- 0.34}$& 0.00$^{+ 0.18}_{- 0.00}$  & 0.28$^{+ 0.23}_{- 0.17}$ & 0.69$^{+ 0.18}_{- 0.16}$ & 0.27$^{+ 0.24}_{- 0.22}$  & 0.69$^{+ 0.44}_{- 0.37}$& 0.00$^{+ 0.17}_{- 0.00}$\\
Fe  & 0.46$^{+ 0.08}_{- 0.06}$ & 0.31$^{+ 0.05}_{- 0.04}$ & 0.47$^{+ 0.08}_{- 0.07}$  & 0.54$^{+ 0.16}_{- 0.12}$& 0.43$^{+ 0.11}_{- 0.07}$  & 0.43$^{+ 0.08}_{- 0.04}$ & 0.32$^{+ 0.05}_{- 0.05}$ & 0.46$^{+ 0.08}_{- 0.06}$  & 0.52$^{+ 0.13}_{- 0.11}$& 0.45$^{+ 0.09}_{- 0.08}$\\
kT1  & 0.26$^{+ 0.01}_{- 0.01}$ & 0.29$^{+ 0.03}_{- 0.02}$ & 0.26$^{+ 0.01}_{- 0.01}$ & 0.27$^{+ 0.04}_{- 0.02}$& 0.24$^{+ 0.02}_{- 0.02}$ & -- & --  & -- & -- & -- \\
EM1  & 1.62$^{+ 0.32}_{- 0.28}$ & 2.81$^{+ 0.81}_{- 0.53}$ & 1.81$^{+ 0.41}_{- 0.39}$ & 1.63$^{+ 0.81}_{- 0.48}$& 2.06$^{+ 0.56}_{- 0.54}$ & 0.52$^{+ 0.15}_{- 0.15}$ & 0.17$^{+ 0.37}_{- 0.12}$ & 0.56$^{+ 0.21}_{- 0.20}$ & 0.30$^{+ 0.22}_{- 0.23}$& 0.96$^{+ 0.37}_{- 0.36}$\\
kT2 & 0.63$^{+ 0.02}_{- 0.02}$ & 0.60$^{+ 0.03}_{- 0.02}$ & 0.64$^{+ 0.02}_{- 0.02}$  & 0.64$^{+ 0.10}_{- 0.04}$& 0.64$^{+ 0.03}_{- 0.03}$ & -- & --  & -- & -- & -- \\
EM2 & 2.18$^{+ 0.36}_{- 0.31}$ & 3.26$^{+ 0.63}_{- 0.66}$ & 2.57$^{+ 0.54}_{- 0.48}$  & 1.70$^{+ 0.68}_{- 0.70}$& 3.33$^{+ 0.80}_{- 0.82}$ & 0.99$^{+ 0.26}_{- 0.29}$ & 2.88$^{+ 0.63}_{- 0.69}$ & 1.37$^{+ 0.51}_{- 0.46}$  & 1.31$^{+ 0.62}_{- 0.59}$& 1.37$^{+ 0.70}_{- 0.65}$\\
kT3  & 1.20$^{+ 0.14}_{- 0.15}$ & 1.53$^{+ 0.16}_{- 0.12}$ & 1.43$^{+ 0.14}_{- 0.12}$ & 1.35$^{+ 0.13}_{- 0.15}$& 1.64$^{+ 0.28}_{- 0.21}$ & -- & --  & -- & -- & -- \\
EM3 & 0.96$^{+ 0.18}_{- 0.19}$ & 1.02$^{+ 0.17}_{- 0.18}$ & 0.99$^{+ 0.17}_{- 0.18}$  & 0.94$^{+ 0.23}_{- 0.22}$& 1.01$^{+ 0.28}_{- 0.28}$ & 2.14$^{+ 0.29}_{- 0.34}$ & 2.82$^{+ 0.50}_{- 0.45}$ & 2.31$^{+ 0.49}_{- 0.45}$  & 1.70$^{+ 0.64}_{- 0.49}$& 2.74$^{+ 0.78}_{- 0.64}$\\
EM4  & -- & --  & -- & -- & -- & 1.14$^{+ 0.15}_{- 0.20}$& 0.92$^{+ 0.23}_{- 0.29}$& 0.90$^{+ 0.21}_{- 0.20}$ & 0.92$^{+ 0.25}_{- 0.22}$& 0.77$^{+ 0.32}_{- 0.39}$\\
EM5 & -- & --  & -- & -- & -- & 0.00$^{+ 0.20}_{- 0.00}$& 0.27$^{+ 0.10}_{- 0.22}$& 0.23$^{+ 0.11}_{- 0.15}$ & 0.11$^{+ 0.13}_{- 0.11}$& 0.44$^{+ 0.18}_{- 0.24}$\\
EM6 & -- & --  & -- & -- & -- & 0.00$^{+ 0.22}_{- 0.00}$& 0.00$^{+ 0.10}_{- 0.00}$& 0.00$^{+ 0.04}_{- 0.00}$ & 0.00$^{+ 0.07}_{- 0.00}$& 0.00$^{+ 0.10}_{- 0.00}$\\\hline
red.$\chi^2$& 1.26 (731)   & 0.99 (291)     & 1.17 (365) & 0.97 (287)  & 1.06 (303) & 1.29 (731)   & 0.99 (291)     & 1.22 (365) & 1.00 (287)  & 1.10 (303)\\\hline
$L_{\rm X}$& 5.83 & 7.25  & 6.59 & 5.87 & 7.30 & 5.86 & 7.27  & 6.61 & 5.88 & 7.33\\\hline\hline
 & \multicolumn{5}{c}{AT Mic}\\\hline
C  & 0.80$^{+ 0.13}_{- 0.11}$ & -- & -- & -- & -- & 0.79$^{+ 0.13}_{- 0.11}$ & -- & -- & -- & -\\
N  & 0.65$^{+ 0.13}_{- 0.11}$ & -- & -- & -- &--  & 0.66$^{+ 0.13}_{- 0.11}$ & -- & -- & -- &--\\
O  & 0.58$^{+ 0.07}_{- 0.06}$ & 0.37$^{+ 0.04}_{- 0.03}$ & 0.45$^{+ 0.06}_{- 0.04}$ & 0.44$^{+ 0.05}_{- 0.04}$& 0.46$^{+ 0.09}_{- 0.07}$   & 0.62$^{+ 0.08}_{- 0.06}$ & 0.37$^{+ 0.04}_{- 0.03}$ & 0.43$^{+ 0.03}_{- 0.03}$ & 0.43$^{+ 0.05}_{- 0.04}$& 0.45$^{+ 0.06}_{- 0.06}$\\
Ne  & 1.31$^{+ 0.16}_{- 0.14}$ & 0.83$^{+ 0.08}_{- 0.10}$ & 0.67$^{+ 0.07}_{- 0.09}$ & 0.63$^{+ 0.09}_{- 0.08}$& 0.72$^{+ 0.24}_{- 0.16}$  & 1.26$^{+ 0.15}_{- 0.14}$ & 0.85$^{+ 0.08}_{- 0.06}$ & 0.78$^{+ 0.05}_{- 0.05}$ & 0.73$^{+ 0.08}_{- 0.07}$& 0.74$^{+ 0.13}_{- 0.11}$ \\
Mg  & 0.30$^{+ 0.10}_{- 0.09}$ & 0.24$^{+ 0.05}_{- 0.05}$ & 0.14$^{+ 0.04}_{- 0.04}$ & 0.14$^{+ 0.06}_{- 0.05}$& 0.13$^{+ 0.09}_{- 0.08}$  & 0.27$^{+ 0.09}_{- 0.09}$ & 0.28$^{+ 0.06}_{- 0.06}$ & 0.20$^{+ 0.04}_{- 0.04}$ & 0.20$^{+ 0.05}_{- 0.06}$& 0.20$^{+ 0.10}_{- 0.09}$ \\
Si  & 0.40$^{+ 0.19}_{- 0.19}$ & 0.49$^{+ 0.09}_{- 0.08}$ & 0.25$^{+ 0.05}_{- 0.04}$ & 0.28$^{+ 0.07}_{- 0.06}$& 0.21$^{+ 0.10}_{- 0.09}$  & 0.39$^{+ 0.19}_{- 0.19}$ & 0.47$^{+ 0.08}_{- 0.07}$ & 0.31$^{+ 0.05}_{- 0.05}$ & 0.37$^{+ 0.07}_{- 0.07}$& 0.28$^{+ 0.11}_{- 0.10}$ \\
S  & 0.17$^{+ 0.11}_{- 0.10}$ & 0.44$^{+ 0.14}_{- 0.14}$ & 0.27$^{+ 0.11}_{- 0.11}$ & 0.27$^{+ 0.15}_{- 0.14}$& 0.11$^{+ 0.24}_{- 0.11}$   & 0.18$^{+ 0.12}_{- 0.11}$ & 0.50$^{+ 0.14}_{- 0.13}$ & 0.22$^{+ 0.11}_{- 0.10}$ & 0.25$^{+ 0.18}_{- 0.15}$& 0.16$^{+ 0.20}_{- 0.16}$\\
Fe  & 0.35$^{+ 0.05}_{- 0.04}$ & 0.18$^{+ 0.03}_{- 0.02}$ & 0.21$^{+ 0.02}_{- 0.02}$ & 0.23$^{+ 0.07}_{- 0.08}$& 0.20$^{+ 0.06}_{- 0.03}$   & 0.32$^{+ 0.04}_{- 0.04}$ & 0.23$^{+ 0.03}_{- 0.03}$ & 0.23$^{+ 0.02}_{- 0.02}$ & 0.24$^{+ 0.03}_{- 0.03}$& 0.27$^{+ 0.06}_{- 0.04}$\\
kT1  & 0.27$^{+ 0.01}_{- 0.01}$ & 0.36$^{+ 0.04}_{- 0.03}$ & 0.25$^{+ 0.01}_{- 0.01}$ & 0.26$^{+ 0.01}_{- 0.01}$& 0.24$^{+ 0.02}_{- 0.03}$  & -- & --  & -- & -- & -- \\
EM1  & 4.97$^{+ 0.79}_{- 0.69}$ & 11.20$^{+ 5.64}_{- 2.39}$ & 7.42$^{+ 1.01}_{- 1.60}$ & 8.20$^{+ 1.33}_{- 1.23}$& 6.50$^{+ 2.30}_{- 3.23}$   & 0.73$^{+ 0.31}_{- 0.26}$ & 0.00$^{+ 0.23}_{- 0.00}$ & 2.61$^{+ 0.54}_{- 0.49}$ & 2.40$^{+ 0.75}_{- 0.65}$& 3.06$^{+ 1.11}_{- 1.10}$\\
kT2 & 0.63$^{+ 0.02}_{- 0.01}$ & 0.60$^{+ 0.18}_{- 0.04}$ & 0.64$^{+ 0.02}_{- 0.02}$ & 0.65$^{+ 0.02}_{- 0.02}$& 0.62$^{+ 0.03}_{- 0.03}$ & -- & --  & -- & -- & -- \\
EM2 & 7.34$^{+ 1.01}_{- 0.93}$ & 9.19$^{+ 3.26}_{- 5.42}$ & 16.01$^{+ 2.46}_{- 1.49}$& 13.91$^{+ 1.66}_{- 1.81}$& 19.15$^{+ 4.75}_{- 6.17}$ & 4.06$^{+ 0.80}_{- 0.78}$ & 8.46$^{+ 1.6}_{- 1.07}$ & 6.58$^{+ 0.96}_{- 1.11}$& 6.92$^{+ 1.66}_{- 1.81}$& 6.72$^{+ 2.75}_{- 2.56}$\\
kT3  & 1.98$^{+ 0.39}_{- 0.28}$ & 2.06$^{+ 0.27}_{- 0.12}$ & 2.56$^{+ 0.48}_{- 0.19}$ & 2.04$^{+ 1.34}_{- 1.84}$& 3.00$^{+ 0.04}_{- 0.03}$ & -- & --  & -- & -- & -- \\
EM3 & 6.55$^{+ 0.66}_{- 0.69}$ & 7.93$^{+ 0.71}_{- 1.08}$ & 5.52$^{+ 0.65}_{- 1.09}$ & 2.39$^{+ 0.74}_{- 0.59}$& 11.86$^{+ 2.78}_{- 2.18}$  & 7.28$^{+ 1.12}_{- 0.96}$ & 10.16$^{+ 1.94}_{- 1.68}$ & 12.71$^{+ 1.60}_{- 0.84}$ & 11.51$^{+ 2.15}_{- 1.33}$& 11.12$^{+ 3.19}_{- 2.93}$\\
EM4  & -- & --  & -- & -- & -- & 1.78$^{+ 1.18}_{- 1.15}$& 4.57$^{+ 1.13}_{- 1.79}$& 0.84$^{+ 1.00}_{- 0.84}$ & 1.49$^{+ 0.72}_{- 0.72}$& 4.98$^{+ 2.11}_{- 2.14}$\\
EM5 & -- & --  & -- & -- & -- & 4.02$^{+ 2.39}_{- 4.02}$& 2.24$^{+ 2.16}_{- 1.28}$& 5.74$^{+ 0.99}_{- 0.59}$ & 1.85$^{+ 0.42}_{- 0.32}$& 6.27$^{+ 3.19}_{- 3.11}$\\
EM6 & -- & --  & -- & -- & -- & 2.39$^{+ 3.17}_{- 2.39}$& 1.62$^{+ 0.60}_{- 0.99}$& 0.21$^{+ 0.47}_{- 0.21}$ & 0.00$^{+ 0.09}_{- 0.00}$& 4.02$^{+ 1.44}_{- 1.48}$\\\hline
red.$\chi^2$& 1.15 (1299) & 1.39 (231)  & 1.31 (537) & 1.14 (424) & 1.10 (511)& 1.16 (1299) & 1.38 (231)  & 1.34 (537) & 1.23 (424) & 1.06 (511) \\\hline
$L_{\rm X}$& 22.71 & 27.64  & 28.95 & 23.61 & 39.65& 23.49 & 27.87  & 29.01 & 23.68 & 39.75\\\hline
\end{tabular}
}
\end{table*}

\addtocounter{table}{-1}
\begin{table*}[!ht]
\setlength\tabcolsep{5pt}
\caption{{\bf cont.}}
{\scriptsize
\begin{tabular}{l|ccccc|cccccc}\hline\hline
Param. & RGS & MOS & PN & PN (low) & PN (high) & RGS & MOS & PN & PN (low) & PN (high)\\\hline

 & \multicolumn{5}{c}{AD Leo}\\\hline
C  & 0.83$^{+ 0.10}_{- 0.09}$ & -- & -- & -- & --  & 0.85$^{+ 0.11}_{- 0.10}$ & -- & -- & -- & --\\
N  & 0.73$^{+ 0.11}_{- 0.10}$ & -- & -- & -- & -- & 0.76$^{+ 0.12}_{- 0.10}$ & -- & -- & -- & -\\
O  & 0.52$^{+ 0.05}_{- 0.05}$ & 0.41$^{+ 0.04}_{- 0.03}$ & 0.45$^{+ 0.04}_{- 0.03}$ & 0.44$^{+ 0.06}_{- 0.05}$& 0.44$^{+ 0.06}_{- 0.04}$  & 0.58$^{+ 0.06}_{- 0.05}$ & 0.41$^{+ 0.04}_{- 0.03}$ & 0.46$^{+ 0.04}_{- 0.04}$ & 0.47$^{+ 0.06}_{- 0.06}$& 0.44$^{+ 0.05}_{- 0.04}$\\
Ne  & 1.08$^{+ 0.11}_{- 0.10}$ & 0.86$^{+ 0.11}_{- 0.08}$ & 0.68$^{+ 0.07}_{- 0.07}$ & 0.68$^{+ 0.10}_{- 0.11}$& 0.71$^{+ 0.10}_{- 0.10}$  & 1.07$^{+ 0.12}_{- 0.10}$ & 0.83$^{+ 0.06}_{- 0.05}$ & 0.74$^{+ 0.06}_{- 0.06}$ & 0.71$^{+ 0.10}_{- 0.08}$& 0.75$^{+ 0.08}_{- 0.07}$\\
Mg  & 0.44$^{+ 0.09}_{- 0.09}$ & 0.30$^{+ 0.06}_{- 0.05}$ & 0.14$^{+ 0.04}_{- 0.04}$ & 0.14$^{+ 0.05}_{- 0.05}$& 0.13$^{+ 0.06}_{- 0.05}$  & 0.41$^{+ 0.08}_{- 0.08}$ & 0.30$^{+ 0.05}_{- 0.05}$ & 0.17$^{+ 0.04}_{- 0.04}$ & 0.15$^{+ 0.06}_{- 0.04}$& 0.17$^{+ 0.07}_{- 0.05}$\\
Si  & 0.56$^{+ 0.17}_{- 0.16}$ & 0.67$^{+ 0.08}_{- 0.07}$ & 0.40$^{+ 0.06}_{- 0.05}$& 0.40$^{+ 0.08}_{- 0.08}$ & 0.41$^{+ 0.09}_{- 0.08}$  & 0.56$^{+ 0.17}_{- 0.17}$ & 0.67$^{+ 0.07}_{- 0.07}$ & 0.45$^{+ 0.06}_{- 0.06}$& 0.42$^{+ 0.09}_{- 0.07}$ & 0.46$^{+ 0.09}_{- 0.07}$\\
S  & 0.41$^{+ 0.10}_{- 0.09}$ & 0.61$^{+ 0.11}_{- 0.12}$ & 0.36$^{+ 0.12}_{- 0.12}$ & 0.40$^{+ 0.21}_{- 0.18}$& 0.26$^{+ 0.17}_{- 0.13}$  & 0.46$^{+ 0.12}_{- 0.11}$ & 0.63$^{+ 0.12}_{- 0.12}$ & 0.36$^{+ 0.14}_{- 0.13}$ & 0.43$^{+ 0.22}_{- 0.20}$& 0.26$^{+ 0.17}_{- 0.14}$\\
Fe  & 0.47$^{+ 0.04}_{- 0.04}$ & 0.30$^{+ 0.03}_{- 0.03}$ & 0.27$^{+ 0.03}_{- 0.02}$ & 0.29$^{+ 0.03}_{- 0.03}$& 0.27$^{+ 0.03}_{- 0.03}$  & 0.44$^{+ 0.05}_{- 0.04}$ & 0.31$^{+ 0.03}_{- 0.02}$ & 0.28$^{+ 0.02}_{- 0.02}$ & 0.29$^{+ 0.03}_{- 0.03}$& 0.29$^{+ 0.04}_{- 0.03}$\\
kT1  & 0.26$^{+ 0.01}_{- 0.01}$ & 0.31$^{+ 0.02}_{- 0.03}$ & 0.24$^{+ 0.01}_{- 0.01}$ & 0.24$^{+ 0.01}_{- 0.01}$& 0.25$^{+ 0.01}_{- 0.01}$& -- & --  & -- & -- & --\\
EM1  & 1.23$^{+ 0.15}_{- 0.14}$ & 1.82$^{+ 0.28}_{- 0.32}$ & 1.63$^{+ 0.21}_{- 0.21}$ & 1.75$^{+ 0.27}_{- 0.30}$& 1.69$^{+ 0.29}_{- 0.39}$  & 0.29$^{+ 0.12}_{- 0.10}$ & 0.00$^{+ 0.02}_{- 0.00}$ & 0.78$^{+ 0.11}_{- 0.11}$ & 0.83$^{+ 0.14}_{- 0.14}$& 0.71$^{+ 0.16}_{- 0.17}$\\
kT2 & 0.63$^{+ 0.02}_{- 0.02}$ & 0.61$^{+ 0.02}_{- 0.02}$ & 0.62$^{+ 0.01}_{- 0.01}$ & 0.60$^{+ 0.02}_{- 0.02}$& 0.62$^{+ 0.02}_{- 0.02}$& -- & --  & -- & -- & --\\
EM2 & 1.41$^{+ 0.15}_{- 0.13}$ & 2.06$^{+ 0.27}_{- 0.29}$ & 3.22$^{+ 0.30}_{- 0.29}$ & 2.66$^{+ 0.35}_{- 0.27}$& 3.71$^{+ 0.62}_{- 0.48}$ & 0.85$^{+ 0.13}_{- 0.13}$ & 1.73$^{+ 0.15}_{- 0.19}$ & 0.98$^{+ 0.29}_{- 0.23}$ & 1.00$^{+ 0.35}_{- 0.27}$& 1.19$^{+ 0.46}_{- 0.36}$\\
kT3  & 1.33$^{+ 0.15}_{- 0.10}$ & 1.54$^{+ 0.09}_{- 0.08}$ & 2.18$^{+ 0.50}_{- 0.29}$ & 2.65$^{+ 2.64}_{- 0.64}$& 1.96$^{+ 0.50}_{- 0.23}$& -- & --  & -- & -- & --\\
EM3 & 0.67$^{+ 0.08}_{- 0.09}$ & 0.82$^{+ 0.08}_{- 0.09}$ & 0.43$^{+ 0.10}_{- 0.10}$ & 0.22$^{+ 0.10}_{- 0.010}$& 0.77$^{+ 0.17}_{- 0.23}$ & 1.40$^{+ 0.17}_{- 0.14}$ & 2.02$^{+ 0.26}_{- 0.24}$ & 2.97$^{+ 0.18}_{- 0.38}$ & 2.50$^{+ 0.30}_{- 0.31}$& 3.19$^{+ 0.36}_{- 0.55}$\\
EM4  & -- & --  & -- & -- & -- & 0.64$^{+ 0.12}_{- 0.16}$& 0.69$^{+ 0.12}_{- 0.13}$& 0.08$^{+ 0.21}_{- 0.08}$ & 0.00$^{+ 0.22}_{- 0.00}$& 0.46$^{+ 0.30}_{- 0.21}$\\
EM5 & -- & --  & -- & -- & -- & 0.00$^{+ 0.27}_{- 0.00}$& 0.23$^{+ 0.05}_{- 0.09}$& 0.41$^{+ 0.06}_{- 0.13}$ & 0.21$^{+ 0.10}_{- 0.15}$& 0.51$^{+ 0.09}_{- 0.29}$\\
EM6 & -- & --  & -- & -- & -- & 0.09$^{+ 0.16}_{- 0.09}$& 0.00$^{+ 0.04}_{- 0.00}$& 0.00$^{+ 0.04}_{- 0.00}$ & 0.00$^{+ 0.09}_{- 0.00}$& 0.00$^{+ 0.12}_{- 0.00}$\\\hline
red.$\chi^2$& 1.29 (1444) & 1.31 (206)  & 1.27 (442) & 1.13 (395) & 1.15 (435)& 1.30 (1444) & 1.31 (206)  & 1.29 (442) & 1.12 (395) & 1.17 (435)\\\hline
$L_{\rm X}$& 4.12 & 4.96  & 5.39 & 4.71 & 6.28 & 4.13 & 4.97  & 5.41 & 4.71 & 6.30\\\hline\hline
 & \multicolumn{4}{c}{EV Lac}\\\hline
C  & 0.51$^{+ 0.08}_{- 0.07}$ & -- & --  &  -- & --  & 0.54$^{+ 0.08}_{- 0.07}$ & -- & --  &  -- & -- \\
N  & 0.45$^{+ 0.08}_{- 0.08}$ & -- & --  &  -- &  --  & 0.47$^{+ 0.09}_{- 0.08}$ & -- & --  &  -- &  --\\
O  & 0.39$^{+ 0.05}_{- 0.04}$ & 0.31$^{+ 0.03}_{- 0.03}$ & 0.34$^{+ 0.03}_{- 0.03}$  & 0.35$^{+ 0.04}_{- 0.03}$& 0.32$^{+ 0.06}_{- 0.04}$ & 0.45$^{+ 0.05}_{- 0.04}$ & 0.31$^{+ 0.03}_{- 0.03}$ & 0.35$^{+ 0.03}_{- 0.02}$  & 0.36$^{+ 0.04}_{- 0.03}$& 0.34$^{+ 0.04}_{- 0.04}$\\
Ne  & 0.69$^{+ 0.09}_{- 0.08}$ & 0.56$^{+ 0.09}_{- 0.08}$ & 0.48$^{+ 0.05}_{- 0.05}$  & 0.50$^{+ 0.06}_{- 0.06}$& 0.49$^{+ 0.09}_{- 0.10}$  & 0.68$^{+ 0.08}_{- 0.08}$ & 0.62$^{+ 0.06}_{- 0.05}$ & 0.51$^{+ 0.04}_{- 0.03}$  & 0.52$^{+ 0.05}_{- 0.05}$& 0.49$^{+ 0.06}_{- 0.06}$\\
Mg  & 0.35$^{+ 0.08}_{- 0.08}$ & 0.26$^{+ 0.06}_{- 0.05}$ & 0.16$^{+ 0.03}_{- 0.03}$  & 0.20$^{+ 0.04}_{- 0.04}$& 0.12$^{+ 0.06}_{- 0.05}$  & 0.31$^{+ 0.08}_{- 0.07}$ & 0.29$^{+ 0.05}_{- 0.05}$ & 0.19$^{+ 0.03}_{- 0.02}$  & 0.22$^{+ 0.04}_{- 0.04}$& 0.13$^{+ 0.06}_{- 0.04}$\\
Si  & 0.63$^{+ 0.17}_{- 0.17}$ & 0.49$^{+ 0.08}_{- 0.07}$ & 0.36$^{+ 0.05}_{- 0.04}$  & 0.42$^{+ 0.06}_{- 0.06}$& 0.30$^{+ 0.07}_{- 0.07}$  & 0.62$^{+ 0.17}_{- 0.17}$ & 0.52$^{+ 0.07}_{- 0.07}$ & 0.40$^{+ 0.04}_{- 0.04}$  & 0.45$^{+ 0.06}_{- 0.06}$& 0.32$^{+ 0.08}_{- 0.05}$\\
S  & 0.18$^{+ 0.09}_{- 0.08}$ & 0.48$^{+ 0.13}_{- 0.12}$ & 0.32$^{+ 0.10}_{- 0.09}$  & 0.40$^{+ 0.13}_{- 0.13}$& 0.18$^{+ 0.14}_{- 0.13}$  & 0.21$^{+ 0.10}_{- 0.09}$ & 0.51$^{+ 0.13}_{- 0.12}$ & 0.33$^{+ 0.10}_{- 0.09}$  & 0.41$^{+ 0.15}_{- 0.12}$& 0.19$^{+ 0.15}_{- 0.12}$\\
Fe  & 0.38$^{+ 0.04}_{- 0.04}$ & 0.24$^{+ 0.03}_{- 0.03}$ & 0.24$^{+ 0.02}_{- 0.02}$  & 0.27$^{+ 0.02}_{- 0.02}$& 0.22$^{+ 0.03}_{- 0.03}$  & 0.35$^{+ 0.04}_{- 0.03}$ & 0.26$^{+ 0.03}_{- 0.03}$ & 0.26$^{+ 0.02}_{- 0.02}$  & 0.28$^{+ 0.02}_{- 0.03}$& 0.23$^{+ 0.04}_{- 0.02}$\\
kT1  & 0.27$^{+ 0.01}_{- 0.01}$ & 0.33$^{+ 0.04}_{- 0.03}$ & 0.25$^{+ 0.02}_{- 0.02}$ & 0.26$^{+ 0.01}_{- 0.01}$& 0.25$^{+ 0.01}_{- 0.02}$& -- & --  & -- & -- & --\\
EM1  & 1.42$^{+ 0.27}_{- 0.21}$ & 2.41$^{+ 0.95}_{- 0.48}$ & 2.01$^{+ 0.27}_{- 0.24}$ & 2.07$^{+ 0.29}_{- 0.27}$& 2.06$^{+ 0.51}_{- 0.69}$  & 0.23$^{+ 0.08}_{- 0.08}$ & 0.00$^{+ 0.09}_{- 0.00}$ & 0.72$^{+ 0.12}_{- 0.12}$ & 0.63$^{+ 0.13}_{- 0.13}$& 0.92$^{+ 0.23}_{- 0.24}$\\
kT2 & 0.63$^{+ 0.02}_{- 0.01}$ & 0.64$^{+ 0.07}_{- 0.03}$ & 0.61$^{+ 0.01}_{- 0.01}$  & 0.61$^{+ 0.01}_{- 0.01}$& 0.61$^{+ 0.02}_{- 0.02}$& -- & --  & -- & -- & --\\
EM2 & 2.04$^{+ 0.24}_{- 0.25}$ & 2.74$^{+ 0.46}_{- 0.87}$ & 3.62$^{+ 0.30}_{- 0.34}$  & 3.10$^{+ 0.31}_{- 0.31}$& 4.41$^{+ 0.95}_{- 0.68}$ & 0.99$^{+ 0.21}_{- 0.22}$ & 2.00$^{+ 0.26}_{- 0.27}$ & 1.54$^{+ 0.30}_{- 0.32}$  & 1.69$^{+ 0.35}_{- 0.38}$& 1.28$^{+ 0.70}_{- 0.37}$\\
kT3  & 1.62$^{+ 0.55}_{- 0.27}$ & 1.85$^{+ 0.24}_{- 0.17}$ & 2.18$^{+ 0.30}_{- 0.23}$ & 2.24$^{+ 0.74}_{- 0.39}$& 2.05$^{+ 0.48}_{- 0.21}$& -- & --  & -- & -- & --\\
EM3 & 0.60$^{+ 0.14}_{- 0.14}$ & 0.72$^{+ 0.12}_{- 0.13}$ & 0.53$^{+ 0.11}_{- 0.09}$  & 0.28$^{+ 0.10}_{- 0.09}$& 1.11$^{+ 0.23}_{- 0.32}$ & 2.11$^{+ 0.25}_{- 0.23}$ & 2.85$^{+ 0.44}_{- 0.38}$ & 3.22$^{+ 0.30}_{- 0.32}$  & 2.74$^{+ 0.41}_{- 0.34}$& 4.07$^{+ 0.61}_{- 0.82}$\\
EM4  & -- & --  & -- & -- & -- & 0.42$^{+ 0.17}_{- 0.22}$& 0.45$^{+ 0.20}_{- 0.21}$& 0.20$^{+ 0.14}_{- 0.17}$ & 0.12$^{+ 0.15}_{- 0.12}$& 0.43$^{+ 0.43}_{- 0.15}$\\
EM5 & -- & --  & -- & -- & -- & 0.00$^{+ 0.49}_{- 0.00}$& 0.40$^{+ 0.08}_{- 0.19}$& 0.44$^{+ 0.06}_{- 0.10}$ & 0.24$^{+ 0.06}_{- 0.12}$& 0.81$^{+ 0.11}_{- 0.05}$\\
EM6 & -- & --  & -- & -- & -- & 0.26$^{+ 0.19}_{- 0.26}$& 0.00$^{+ 0.08}_{- 0.00}$& 0.00$^{+ 0.04}_{- 0.00}$ & 0.00$^{+ 0.04}_{- 0.00}$& 0.00$^{+ 0.15}_{- 0.00}$\\\hline
red.$\chi^2$& 1.11 (1173) & 1.35 (196)  & 1.32 (479) & 1.15 (459) & 1.19 (457) & 1.12 (1173) & 1.40 (196)  & 1.33 (479) & 1.15 (459) & 1.20 (457)\\\hline
$L_{\rm X}$& 4.26 & 5.30  & 5.53 & 4.96 & 6.66 & 4.37 &5.31  & 5.54 & 4.97 & 6.68\\\hline
\end{tabular}
}
\end{table*}

Closer inspection of the M-dwarf spectra and their individual
lines reveals further differences. To quantify these differences the spectra from each type of detector were fitted
separately with the \hbox{3\,-T} model and the \hbox{6\,-T} model. 
Integration times of the spectra may differ due to varying exposure times and the influence of
high background periods on the individual detectors.
We examined the influence of flaring on the temperatures, emission measures and abundances of the plasma.
For this purpose we extracted spectra from flaring and quasi-quiescent intervals as defined in Table~\ref{selflare}
and modelled the PN spectra for the different phases of activity,
keeping all model parameters as free parameters.
The results of this fitting procedure are summarized in Table~\ref{res};
all errors denote 90\,\% confidence range and were calculated by allowing
variations of all other free model parameters. Note the corrected flux for EQ~Peg compared to \cite{eqpeg}.
The quality of these fits in terms of the $\chi^2$ test statistic did not improve when
further components were added to the model.
We point out that the calculated errors are statistical errors only
and additional uncertainties are introduced by the atomic data and instrumental calibration. 

Comparing the fit results for the different instruments one finds
good agreement in general, but deviations of particular results are obvious.
One source of uncertainty is based on the interdependence of emission measure (EM) and elemental abundances; 
a higher EM of a temperature component leads to the
reduction of the abundance of elements producing the strongest lines in this temperature range and vice versa. 
The difference in fit quality between two models is often only marginal. This is especially true for EPIC
data, where the best-fitting models result sometimes in higher EM with lower abundances compared to RGS data. 
However, a combination of these properties is in good agreement, e.g. a simple scaling of the
summed EM with the summed abundances of iron and oxygen (see Table~\ref{res2}). 
Iron and oxygen are the major contributing elements of the emission of M~dwarfs in the considered energy range;
their abundances can therefore be determined best. 
Similar results were obtained when using the oxygen abundance for the cooler components 
and iron for the hotter components.
Also the emission measure is in reality a continuous distribution which results in an 
additional interdependence between temperatures and strength of the respective EM components,
noticeably in larger, often dependent, errors affecting neighbouring components. 
The \hbox{3\,-T} and the \hbox{6\,-T} model lead to fully-consistent results on abundances for all targets;
differences are within the errors and negligible when compared
to the uncertainties arising from the use of different detectors.

\begin{table}[!ht]
\caption{\label{res2}Deviations of the XMM instruments from the mean scaled 
summed EM ($\Sigma\,EM * (A_{Fe}+A_{0}$)[$10^{51}\,$cm$^{-3}$]).}
\begin{center}
{\scriptsize
\begin{tabular}{l|cccc}\hline\hline
Instr. & EQ Peg & AT Mic & AD Leo & EV Lac \\\hline
RGS & -5\,\% & +3\,\% & -4\,\% & -5\,\% \\
PN  & +10\,\%  & +7\,\%  & +9\,\% & +7\,\% \\
MOS & -5\,\% & -11\,\% & -4\,\% & -3\,\% \\\hline
sc.\,$\Sigma$\,EM & 4.51 & 15.32 & 3.08 & 2.97\\\hline
\end{tabular}
}
\end{center}
\end{table}

One effect of calibration uncertainties is that the best fit models of the different detectors 
do vary in X-ray flux, especially below 1.0\,keV. Typical RGS models result in a $\sim$\,20\% lower
X-ray flux, mainly in the energy range \hbox{0.2\,-\,1.0\,keV} compared to EPIC, 
whereas the PN models are a few percent more luminous than the MOS models.  
The RGS luminosities in Table\,\ref{res} have been extrapolated 
to the \hbox{0.2\,-\,10.0\,keV} band for comparison with the EPIC luminosities.
Smaller cross-calibration errors, particularly at energies below 0.5\,keV, are also seen in observations made
with the same detector but different setups, especially with different filters.
Enclosed energy correction from annular extraction regions, necessary for piled-up sources, also influences the measured
flux, which can be reduced by up to $\sim$10\,\%. 
Additional effects are expected for the X-ray flux of the binaries, 
but in our case they are dominated by one component and/or
the separation is small compared to the used extraction region.  
Calibration efforts are still ongoing, see e.g. \cite{kirsch} for updated XMM-Newton (cross)-calibration.

We note that that the EPIC detectors are not very appropriate to measure changes in the abundance pattern for
the different phases of activity as they sometimes result in unphysical solutions like a zero abundance. 
On the other hand, the RGS is not very sensitive to high temperature plasma and
in the \hbox{6\,-T} models applied to RGS data sometimes plasma is put into the hottest components,
that is not seen in the EPIC models. Since the contributions from these components in the RGS 
are also compatible with a zero value, we expect the EPIC models to give a more realistic EMD at these energies.
Overall we consider the abundance determination below $\sim$\,1.5\,keV to be more accurate for RGS data, 
while properties of the hotter plasma are best determined with EPIC data. 
Therefore we analysed additionally for the different phases of activity the 
distribution of elemental abundances with RGS data, 
and the emission measures with PN data using a fixed abundance pattern.
The results of flare-related changes will be presented in section \ref{flare}.

The derived temperature structure is consistent among the detectors, independent of the model used,
when accounting for the above mentioned  uncertainties.
The temperature distributions of all sample stars peak consistently
around 7\,-\,8\,MK (1\,keV\,$\equiv$\,11.6\,MK) with the bulk of the plasma being located between 2\,-\,20\,MK
for AD~Leo, EQ~Peg and EV~Lac, whereas the AT~Mic distribution extends to somewhat higher temperatures.
In the EPIC \hbox{3\,-T} models the high temperature component tends to be hotter and/or stronger compared to RGS.

\begin{figure}[!ht]
\includegraphics[width=90mm,]{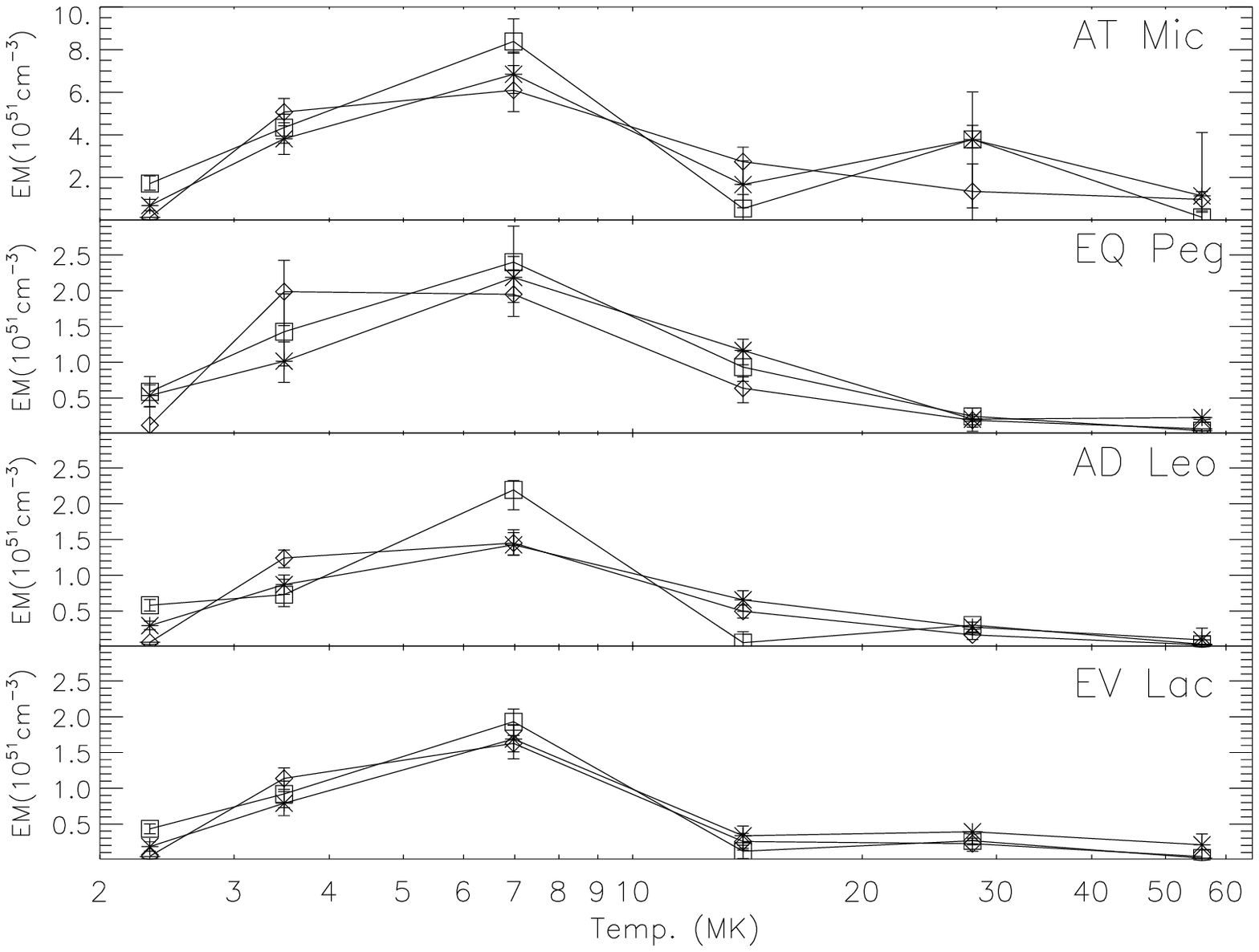}
\caption{\label{emdet}EMDs as derived with the 6-T model and scaled with the summed abundance of iron and oxygen, 
total observation each (RGS: Asterisks, MOS: Diamonds, PN: Squares).}
\end{figure}

The EMDs for the total observation of our sample stars as derived from the \hbox{6\,-T} model with the
different detectors onboard XMM-Newton are shown in Fig.\,\ref{emdet}. Undetected components
are plotted with their upper limits.
The EMDs were scaled with the summed abundance of iron and oxygen and are in good agreement when comparing
the individual instruments but individual temperature components are sometimes more pronounced.  
The shape of the distribution looks similar for our
sample stars but the contributions of the hot plasma differ and are
strongest for AT~Mic, which is also the X-ray brightest source in our sample.
AT~Mic is associated with the $\beta$ Pictoris moving group \citep{barr99,zuck01},
with a probable age of about 10-20 Myr whereas the other stars are members of the kinematic young disk population 
with ages of the order of 1\,Gyr.
At this age AT~Mic would be a still contracting pre-main sequence system, therefore the stars are larger with
higher $L_{\rm bol}$ and because they are in the saturated regime, they have consequently higher $L_{\rm X}$. 

We also investigated joint fits, where data of different instruments, e.g. RGS and MOS, are fitted simultaneously.
While combined fits with datasets of different detectors usually turn out a weighted, intermediate result, 
this is not the case
when taking RGS data in the full energy range and combining it with MOS data above 0.8 or 1.0\,keV.
This choice is reasonable because the relative energy resolution of the EPIC 
instruments improves towards higher energies, whereas the effective area of the RGS drops above 1.0\,keV.
The derived abundances in these fits increase by $\sim$ 20-30\%, whereas emission measure drops, 
compared to the pure RGS model;
the derived temperatures change only marginally. The effect on individual elements and respectively emission measure
components slightly depends on the chosen overlap and is similarly observed with PN data. Discarding the RGS data
at higher energies, e.g. above 1.5\,keV, has only minor influences on the derived models for our stars.

\subsubsection{Coronal abundances}
\label{abusec}

\begin{figure}[!ht]
\vspace*{-0.2cm}
\hspace*{-0.6cm}
\includegraphics[width=95mm,]{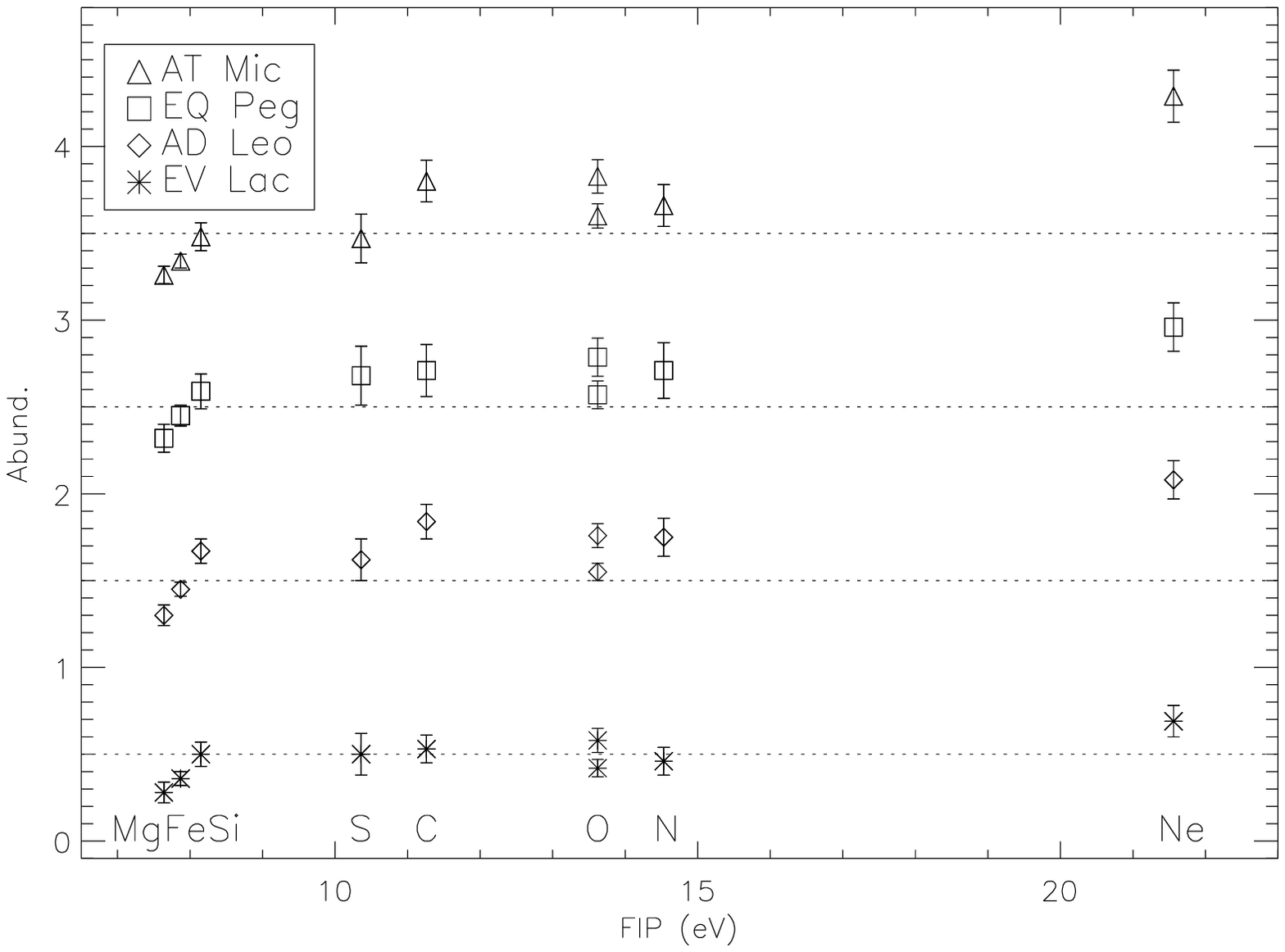}
\caption{\label{fip}Elemental abundances in stellar coronae relative to solar photospheric values 
vs. FIP for our sample stars; an inverse FIP effect is present in all stars. 
Each star is shifted by one solar abundance on the
vertical axis, the dashed lines indicate 0.5 solar abundance in each case.}
\end{figure}

The abundance patterns of our sample stars are shown in Fig.\,\ref{fip}, where we plot the abundances 
with respect to solar photospheric abundances against the FIP (First Ionization Potential)
of the corresponding element. Here we show the results for the total observation that are constrained best.
The data of the individual stars are separated by an offset of one on the vertical axis and a dotted line is
plotted at 0.5 solar abundance to guide the eye.
For elements with stronger lines in the RGS we accept these values (i.e., for Fe, C, O, N, Ne), 
for the other elements we use the MOS values, which has a better spectral resolution compared to PN. 
The higher oxygen values obtained with the solar oxygen abundance given by \cite{pri01} are also 
plotted, indicating the current uncertainties in the solar abundance pattern. 
A discussion of problems in determining solar photospheric abundances can be found in \cite{hol}.

The determined metallicities of our sample stars vary only moderately between the individual stars;
we do not consider these differences to be very significant. However,
all abundances are subsolar when compared to solar photospheric values.
Unfortunately, photospheric abundances for M~dwarfs are usually unknown and the
rare measurements have large uncertainties.
\cite{naf92} published photospheric abundances for AD~Leo, arguing for
a metallicity of $\sim$\,0.4 solar for the analysed low FIP metals; iron was determined to $\sim$\,0.5 solar,
in agreement with our coronal measurements but chromospheric contamination could not be ruled out completely. 
\cite{rotabu} published a metallicity of $\sim$\,0.6 solar for a number of K and M dwarfs from optical/IR observations.
These abundance values are challenging because our sample stars
are members of the kinematic young disk population (with AT~Mic being even younger), 
that are fast rotating \citep{rot1} and highly X-ray active, which
indicates a younger age in contrast to the reduced metallicity one would associate with older stars.

If the photospheric metallicities in our sample stars are indeed as low as suggested by the
authors cited above, i.e. in the range 0.4\,-\,0.6 solar,
the abundances of low FIP elements would actually be consistent with the stellar photospheric values or
possibly slightly depleted, while the high FIP elements would actually be enhanced in the coronal material.
An overall coronal metal abundance deficiency would then not be present,
rather one would find a situation reminiscent of the solar corona, i.e., an enhancement of
specific (but different) elements.

When comparing star-to-star variations of abundance as a function of FIP, one finds a tendency that the variations of
low FIP elements are smaller than those of high FIP elements,
e.g. neon varies by more than 50\,\% compared to iron where the effect is below 10\,\%. 
Whereas the absolute values differ moderately, the observed abundance pattern looks very similar for all stars.
In general, abundance ratios are more robust than absolute values
because they are nearly independent of the EM.
We already mentioned that while low activity stars show the FIP effect more active stars 
have shown none or an inverse FIP effect (IFIP) \citep{bri01,aud03}.
As a major trend the inverse FIP effect is clearly present in all analysed 
sources but individual elements show occasional
deviations, e.g. carbon, which is mainly determined by the \ion{C}{vi} line at the low energy end of the RGS. 
The IFIP effect is detected with all instruments;
however, the strength of the IFIP or abundances of individual elements do vary.
This can be mainly attributed to instrumental properties, e.g. the often low abundance of sulfur in RGS fits.
Here the detectable lines lie at the 
high energy end of the detector where the effective area is quite small.
The inverse FIP effect is strongest 
for the most luminous source AT~Mic and appears to be correlated with activity. 
Taking the abundance ratio Ne/Mg as an indicator of the strength 
of the IFIP effect (since these elements have the largest difference in FIP), 
we find the largest ratio for AT~Mic, intermediate values
for AD~Leo and EQ~Peg and the lowest ratio is found for EV~Lac 
which also exhibits the lowest absolute abundance values.
The same trend is observed for other ratios like Ne/Fe or O/Fe and likewise 
in the results obtained with the MOS detector.

\subsection{Flare-related changes}
\label{flare}

Significant variations in the abundance pattern were not detected with the EPIC data. 
We also searched for variations with RGS data, which
suggests an enhancement of low FIP elements like iron during strong flares, most prominent in
the large flare on AT~Mic when compared to the quasi-quiescent phase. 
While the elements with intermediate FIP show no clear trend,
the high FIP element neon may also be somewhat enhanced, a trend already found by \cite{atmic}.
We caution, however, that the measurement errors are large and that a constant abundance pattern 
is also acceptable within the 90\,\% confidence range derived from the model. 
This effect is still weaker or undetected for the other sample stars.

\begin{figure}[!ht]
\includegraphics[width=53mm,angle=-90]{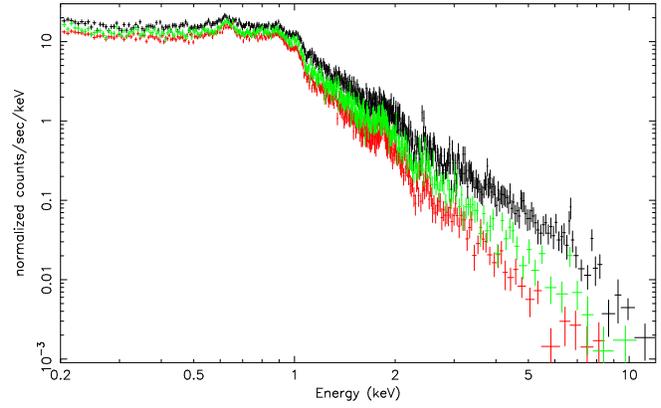}
\caption{\label{atmicspec}Spectra of AT Mic during different states of activity, PN data from selected
time-intervals as defined in Table~\ref{selflare} (high: black, medium: green, qq: red). [{\it Colour figure
in electronic version.}]}
\end{figure}

In contrast to abundances, emission measures and temperatures of the coronal 
plasma change significantly.
As an example we show the PN spectrum of the AT~Mic observation in
Fig.\,\ref{atmicspec}, separated into the three phases of activity.
Not only does the plasma emission measure increase, but
the spectra also harden towards higher activity as more and more emission is found in the higher 
temperature components during flaring. This is also seen in
the growth of the iron lines at 6.7\,keV, which
are predominantly formed at temperatures above 40\,MK.

A quantitative study of the temporal evolution of plasma temperature during flaring was carried out
via the emission measure distribution (EMD).
Because the most significant changes are expected to be in the high 
temperature components of the plasma, we used PN data to study the changes of the EMD. 
For this purpose we use the \hbox{6\,-T} model,
and keep the abundances fixed at the values as determined from the total observation 
to derive the EMDs for the different phases of activity. 
The first two components 
represent the cool plasma (2\,-\,5\,MK), followed by the medium temperature plasma (5\,-\,20\,MK) and the 
last two account for the hot plasma (20\,-\,70\,MK).

\begin{figure}[!ht]
\includegraphics[height=55mm,width=90mm]{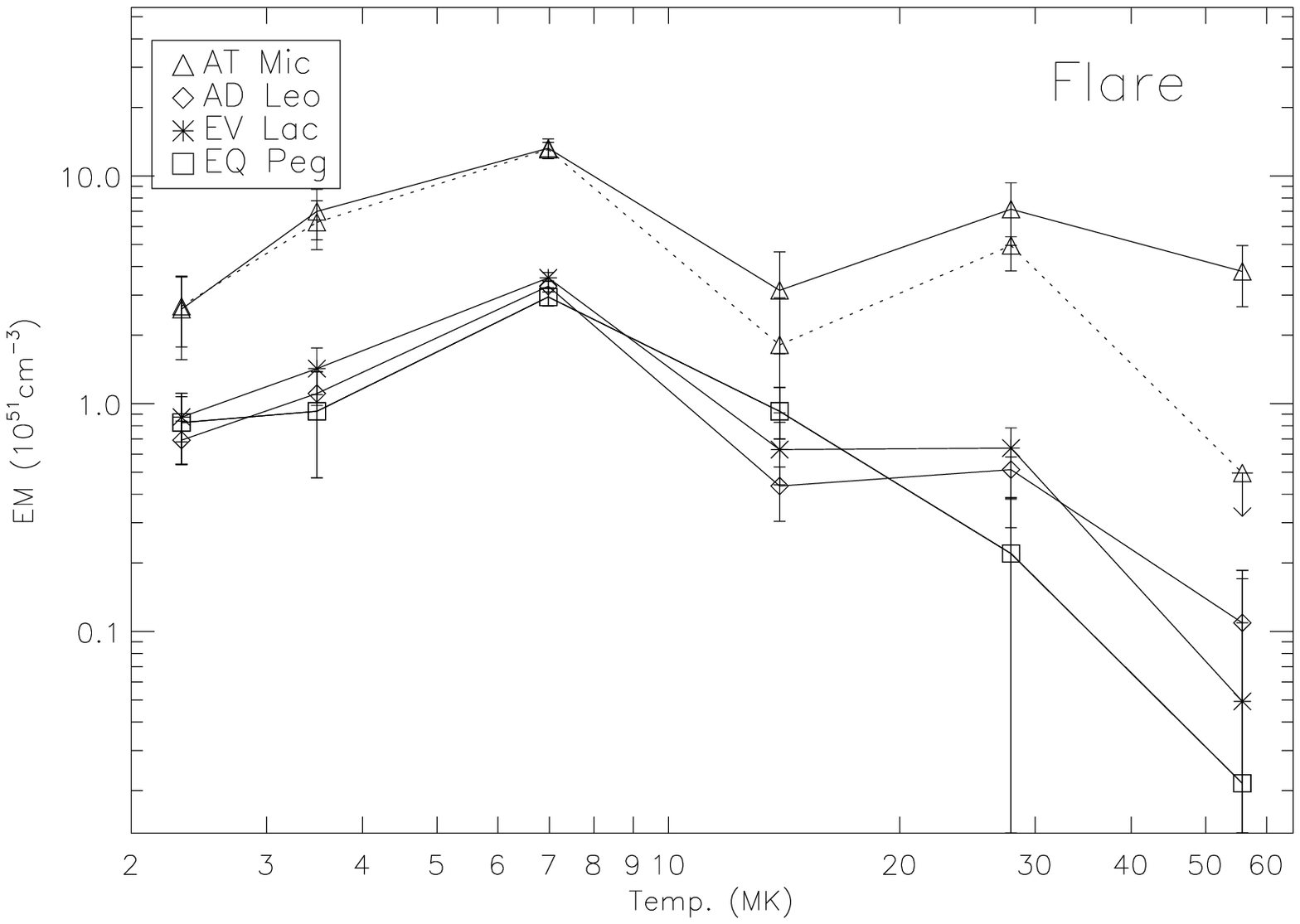}
\includegraphics[height=55mm,width=90mm]{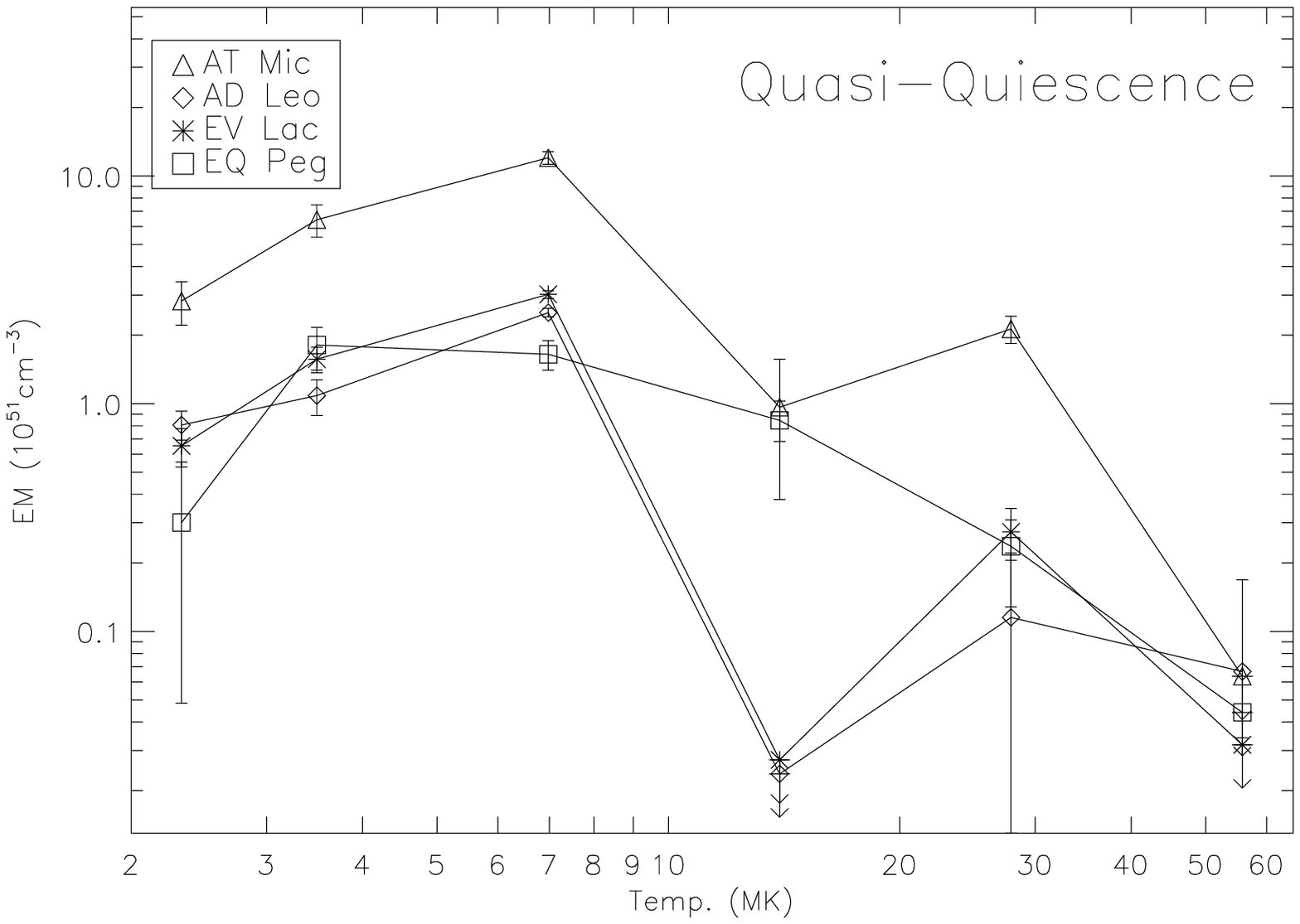}
\caption{\label{3em}EMDs of our sample M~dwarfs in different states of activity derived from PN data.
The upper plot shows the flaring phase, for AT~Mic the medium activity phase is 
additionally plotted as a dashed line while the lower plot shows the quasi-quiescent state.
Undetected components are plotted with their upper limits, denoted by downward arrows.}
\end{figure}

The observed sources vary in terms of X-ray luminosity and its increase during flaring.
The largest increase is found for the most luminous source AT~Mic, where the X-ray luminosity
during the flare peak phase reaches \hbox{$\sim\,44*10^{28}$\,erg\,s$^{-1}$};
the source brightened by nearly a factor of two for about an hour. 
AD~Leo, EQ~Peg and EV~Lac are about half an order of magnitude X-ray fainter than AT~Mic.
A graphical representation of the results on the EMDs of the lower activity stars separated into
two different phases of activity (three for the most active source AT~Mic) is shown in Fig.\,\ref{3em}. 

The weakest flaring activity is found on EQ~Peg, 
with changes in EM affecting mainly the medium temperature component. 
The summed EM of the cool component is constant within errors whereas
the medium component increases significantly;
typical flare temperatures are around 10\,-\,15\,MK. 
The hot component is only marginally detected,contributing only 3\,\% to the EM,
and does not appear to be significantly affected by flaring.
An intermediate case is observed for AD~Leo and EV~Lac; here
the cool components again stay nearly constant, the
medium components increase significantly and the hot component also increases during flaring.
The flare temperatures are in the 15\,-\,40\,MK range.
Remarkably the existence of two more separated parts of the EMDs derived from PN data appears also
in quasi-quiescence and is similarly found on other temperature grids. However, the respective plasma contributes
only a few percent to the total emission measure and the EMDs derived from MOS data are likewise more  
compact in the sense that the plasma appears to be more concentrated at medium temperatures.

The strongest flaring is observed on AT~Mic; here 
the cooler plasma below the main peak at $\sim$\,7\,MKs is nearly unaffected by the flaring.
The main increase is due to the hottest plasma components and large amounts of additional
plasma were detected during flaring at temperatures between 15\,-\,60\,MK.
In the low state a second weaker peak is visible in the EMD between 25 and 30\,MK, 
followed by a steep decline towards 
higher temperatures, indicating no significant amount of plasma above 40\,MK.
With intermediate conditions during the medium activity phase the plasma 
distribution during the large flare exceeds 60\,MK. The contribution of the hot component to the EMD
during the flare is around 25\,\%.
The influences of possible changes of the abundance pattern on the EMD
are minor and we find that the basic behaviour and shape of the EMD is comparable for models
with free or fixed abundances. 

Our analysis of the EMDs shows that flaring does not affect the cool plasma in
a significant way. The shape of the EMDs are quite alike at cooler temperatures for all stars
and peak at temperatures around 7\,-\,8\,MK whereas
the absolute value correlates with X-ray luminosity.
The main influence during flares is found on the amount and temperature of the hotter plasma:
stronger flaring and more active stars produce more and hotter plasma. 
As to the emission of the stars in the quasi-quiescent state a relation between
X-ray luminosity and average plasma temperature is suggested while
the overall shape of the EMDs is again similar. Only the distribution of
binary system EQ~Peg, whose EMD is created by two active and slightly different stars (see \cite{eqpeg}), 
appears broader in all detectors.
A hot tail is quite common, sometimes with
a second smaller peak, extending to $\sim$ 30\,MK in the EMD. This may
indicate the existence of separated plasma components at different temperatures
which are located in different coronal structures even in quasi-quiescence, i.e. 
the main part of the plasma in cooler, more quiescent structures
and a smaller amount in hotter and more active structures. These structures would also be a natural
explanation for the constant flickering observed in the light curves.

\subsection{Loops}

Assuming quasi-static loop-like structures as building blocks for the coronae of M~dwarfs 
the derived EM distributions suggest coronae composed out
of different loop components with different temperatures. Using the RTV scaling law \citep{rtv} 
for simple loops in hydrostatic equilibrium and with constant pressure and cross-section, and the densities
derived by \cite{helike1}, we are able to calculate the size of the loops. In cgs units we obtain:

\smallskip
$n_{e}L=1.3*10^6\,T^{2}_{max},$
\smallskip

\noindent
with the electron density $n_{e}$, the loop top temperature $T_{max}$ and the (semi) loop length $L$.  
The typical electron densities of our sample stars in log($n_e$) as measured 
from the \ion{O}{vii} triplet are around 10.5, and for the \ion{Ne}{ix} triplet in the
range 11.1-11.5, but for \ion{Ne}{ix} the errors are about 0.5 translating into a factor of three in
linear units. The \ion{O}{vii} triplet traces plasma around 2\,MK, the \ion{Ne}{ix} triplet plasma around 4\,MK, thus
indicating denser plasma at higher temperatures. 
However both triplets trace only the relatively cool plasma since
we find average coronal temperatures of about 10\,MK for the less active stars and 15\,MK for AT~Mic 
from the EMDs, in good agreement with the values calculated 
with a scaling law correlating X-ray luminosity and stellar radius ($R_*$)
with average temperature \citep{helike1}. 
While the assumptions of quasi-static loops and the measured densities can be well justified
for the cooler and the quiescent components,
results for the hotter plasma should be taken as upper limits, especially when
taking into account the plausible higher density of the hotter plasma.

Assuming stellar radii of 0.35/0.3 solar radius for spectral type M3.5/M4.5 and
for the binaries for simplicity one dominant component,
we determine loop lengths for the coolest plasma (2\,-\,5\,MK) below 0.01\,$R_*$ on all stars.
Taking the mean plasma density derived from neon, the loop lengths
are about 0.03\,-0.07\,$R_*$ for 
the average coronal temperature of our stars and increase up to 0.3\,$R_*$ for the hotter plasma components.
\cite{helike1} find the density of the hotter plasma consistent with the
low density limit (log\,$n_e$$\approx$12.7) for the used line ratios of \ion{Fe}{xxi}, so an
higher density as the one derived from \ion{Ne}{ix} in the hotter loops would further 
scale down the determined loop length according to the above scaling law;
instead of an increased loop size, an increased density also can explain the observed temperatures.
Repeating the calculations with the low density limit of \ion{Fe}{xxi}, we derive lower limits for the length of the
hotter loops, which are scaled down by a factor of 20 compared to the above derived results.
Therefore we conclude that the dominating coronal structures for the cool plasma are very small, 
i.e.  L\,$<$\,0.01$R_*$, for the medium temperature plasma small
in the sense that L\,$<$\,0.1$R_*$ and still compact (L\,$\la$\,0.3$R_*$) for the hotter plasma
observed here. \cite{favata00a} studied AD~Leo data from various missions, analysed decay times of flare loops 
and likewise found its corona to be compact, even for flare structures.

In this picture the cool emission is produced in relatively small loops nearly unaffected by flaring, 
while medium and hot temperature loops are larger, probably have higher densities and are 
more numerous on more active stars or during flares.

\subsection{Comparison with other results}

We compare our results with other measurements and earlier works to check the consistency of the derived results
and to investigate long-term behaviour. We find that
the results presented here agree well with other analyses, which often used different 
techniques or dealt with individual objects only,
leading to a rather consistent picture of the X-ray properties of active M~dwarfs as a class
and demonstrate the robustness of the results.

Comparison of the measured fluxes of our sample stars with e.g. the ROSAT all-sky survey values \citep{huensch99} 
shows agreement to within a factor less than two. This is of the same order as the variations
found during the moderate activity in the XMM-Newton observations, demonstrating the stability of the X-ray emission.
\cite{favata00a} studied the quasi-quiescent corona of AD~Leo from various missions and found the X-ray
luminosity to be remarkable constant with variations less than a factor of two,
suggesting the absence of strong cyclic variations;
our results agree well, expanding the time period of the observations to over 20 years.
The X-ray luminosity of EQ~Peg determined in this work agrees likewise with the measurements 
made with {\it Einstein} more than 20 years ago.
Combined, these findings establish long term stability of the quasi-quiescent corona for
active M~dwarfs in general.

Comparing our results with previous analysis of the XMM-Newton data,
our AD~Leo results are in good agreement with those obtained by \cite{adleo}, 
who also applied the global fitting approach, 
especially when considering the different spectral code and model, atomic database 
and of course the X-ray dataset used in the analyis.
The same is true for the AT~Mic results presented in \cite{atmic} with a similar analysis;
note that errors stated in these publications are only 1$\sigma$.
Discrepancies concerning the absolute values of emission measure and abundances are 
explained by their mutual interdependence, and a combination, e.g. the product, of these
properties is in much better agreement. 
Concerning the shape of the DEM, which is determined there with polynomial methods from 
MOS and RGS data, results are comparable with our EMD based on multi-temperature fits of PN data.

\cite{maggio} presented a line-based analysis of AD~Leo data taken with {\it Chandra} LETGS 
and also found the inverse FIP effect.
However, they determined the iron abundance from the line-to-continuum ratio in the low-count spectrum
to be almost as high as that in the solar photosphere.
Consequently, nearly all other elements, which are scaled with the iron abundance, appear to be supersolar. 
On the other hand, \cite{favata00a} determined the iron abundance from ASCA data 
to $\sim$\,0.2 solar, low but comparable to our EPIC values.
While the abundances determined with medium resolution spectroscopy are apparently often low, the results 
obtained with the global fits of high resolution data in our work or in \cite{adleo} appear more reliable. 
We also point out that our joint fit approach, which uses a data selection comparable to \cite{aud04}, who
obtained consistent results on the abundances with line-based and global models applied to 
high resolution X-ray spectra,
results in slightly increased but still significant subsolar abundance values. 

Whereas the global metallicities were not determined uniquely in the various analysis,
the inverse FIP effect is consistently found in all analysis of active M~dwarfs, 
despite the fact that different spectral models and atomic codes
were applied to medium and high resolution X-ray data, 
demonstrating that the inverse FIP effect is certainly a robust result.

\section{Summary and Conclusions}
\label{summ}
We have carried out a comparative analysis of XMM-Newton observations of 
M~dwarfs with spectral type \hbox{M3.5\,-\,M4.5}, 
i.\,e., AT~Mic, EQ~Peg, AD~Leo and EV~Lac.
Light curves revealed frequent flaring of all observed sources on different timescales and 
with various strengths of the individual flares. 
While AD~Leo, EQ~Peg and EV~Lac are comparable with respect to activity 
and X-ray brightness, the probably younger system AT~Mic is half an order of magnitude X-ray brighter.
From X-ray spectroscopy, abundance patterns and emission measure distributions were
derived for the sample stars and for different states of activity.
We find that many X-ray quantities, like the coronal temperature structure and the abundance pattern
of the analysed M~dwarfs are very similar despite the fact that
the X-ray luminosity and flaring behaviour differs significantly. 

All analysed M~dwarf coronae show the inverse FIP effect, demonstrated in this work
for the first time for EQ~Peg and EV~Lac, establishing this effect for the coronae of active M~dwarfs as a class.
The determined abundances are lower than solar photospheric abundances;
however, the few available measurements of stellar photospheric metallicities for 
M~dwarfs also point to subsolar values, comparable
with our coronal measurements.
This suggests that a strong coronal metal abundance deficiency may actually
not be present in our sample stars.
Properties like X-ray brightness, temperature of the flare plasma and thus hardness of the X-ray spectrum
are highest for the most active source AT~Mic, and increase during flares for all observed sources, 
suggesting a correlation between these properties and activity.
The strength of the inverse FIP effect is also greatest in AT~Mic, but does not increase during flaring.
Time-resolved abundance analysis suggests that preferably low FIP elements
are enhanced during flares, but this finding is not statistically significant. 

We find from the derived EMDs of the stars that
the plasma temperature structures of the coronae are similar, with a maximum at 7-8\,MK and a hot tail
with a stronger contribution of hotter plasma in the X-ray brighter stars.
The emission measures of the hotter plasma components increase significantly
during flares and higher temperatures are reached in stronger flares, while
the cool plasma is only marginally affected by flaring.
Assuming that the coronae are composed of loop-like structures, this indicates a distribution of structures
in which the cooler plasma is confined in many smaller loops while other, presumably 
larger and denser loops, dominate the additional hot emission generated by flare events. 
The coronae are found to be compact, i.e. the largest loop lengths are smaller than one third stellar radius.

From comparison with older measurements we conclude that the global X-ray properties show long-term stability
while short time-scale behaviour is characterised by frequent variations on top of a level of 'baseline activity'.
With the XMM-Newton data we are able to investigate the X-ray properties of stellar coronae to a previously
unknown precision.
As these observations provide only a snapshot in the life of a few M~dwarfs, larger samples or 
longer observations may show more extreme or different phenomena.  
Nevertheless, the consistency and congruence makes our results appear typical for active M~dwarfs.

\begin{acknowledgements}
This work is based on observations obtained with XMM-Newton, an ESA science
mission with instruments and contributions directly funded by ESA Member
States and the USA (NASA).\\
This research has made use of the SIMBAD database, operated at CDS, Strasbourg, France.
(http://simbad.u-strasbg.fr)\\
We thank the referee for useful comments.\\
J.R. acknowledges support from DLR under 50OR0105.

\end{acknowledgements}


\begin{thebibliography}{}
\bibitem[Allende Prieto et al.(2001)]{pri01}Allende Prieto, C.A., Lambert, D.L., Asplund, M. 2001, ApJ, 556, L63
\bibitem[Anders \& Grevesse(1989)]{and89}Anders E., Grevesse N. 1989, Geo- et Cosmochimica Acta, 53, 197
\bibitem[Arnaud(1996)]{xspec}Arnaud, K.A. 1996, ASP Conf. Series, 101, 17
\bibitem[Audard et al.(2003)]{aud03}Audard, M., G\"udel, M., Sres, A., et al. 2003, A\&A, 398, 1137
\bibitem[Audard et al.(2004)]{aud04}Audard, M., Telleschi, A., G\"udel, M., et al. 2004, ApJ, 617, 531
\bibitem[Barrado y Navascues et al.(1999)]{barr99}Barrado y Navascues, D., Stauffer, J.R., Song, I., Caillault, J.-P. 1999, ApJ 520, L123
\bibitem[Besselaar et al.(2003)]{adleo}van den Besselaar, E.J.M.,Raassen, A.J.J., Mewe, R., et al. 2003,  A\&A, 411, 587
\bibitem[Brinkman et al.(2001)]{bri01}Brinkman, A.C., Behar, E., G\"udel, M., et al. 2001, A\&A, 365, L324
\bibitem[Delfosse et al.(1998)]{rot1}Delfosse, X., Forveille, T., Perrier, C., Mayor, M. 1998, A\&A, 331, 581
\bibitem[Ehle et al.(2003)]{xmm}Ehle, M., Breitfellner, M., Gonzales Riestra, M., et al. 2003, XMM-Newton User's Handbook
\bibitem[Ehle et al.(2004)]{sas}Ehle, M., Pollock, A.M.T., Talavera, A., et al. 2004, User's Guide to XMM-Newton Science Analysis System
\bibitem[Favata et al.(2000a)]{favata00a}Favata, F., Reale, F., Micela, et al. 2000, A\&A, 353, 987
\bibitem[Favata et al.(2000b)]{favata00b}Favata, F., Micela, G., Reale, F. 2000, A\&A, 354, 1021
\bibitem[Feldman et al.(1995)]{feldman}Feldman, U., Seely, J.F., Doschek, G.A., et al. 1995, ApJ, 446, 869
\bibitem[Fuhrmeister et al.(2004)]{rot2}Fuhrmeister, B., Schmitt, J.H.M.M., Wichmann, R. 2004, A\&A 417, 701
\bibitem[Grevesse \& Sauval(1998)]{grev98}Grevesse, N., Sauval, A.J. 1998, Space Sci. Rev., 85, 161
\bibitem[G\"udel et al.(2001)]{ab2}G\"udel, M., Audard, M., Briggs, K., et al. 2001, A\&A, 365, L336
\bibitem[G\"udel et al.(2003)]{flares2}G\"udel, M., Audard, M., Kashyap, V.L., et al. 2003, ApJ, 582, 423
\bibitem[G\"udel et al.(2004)]{proxcen}G\"udel, M., Audard, M., Reale, F., et al. 2004, A\&A, 416, 713
\bibitem[Holweger(2001)]{hol}Holweger, H. 2001, AIPC, 598, 23 
\bibitem[Hudson(1991)]{flares3}Hudson, H.S. 1991, SoPh, 133, 357
\bibitem[H\"unsch et al.(1999)]{huensch99}H\"unsch, M., Schmitt, J.H.M.M., Sterzik, M.F., et al. 1999, A\&A Sup. Ser., 135, 319
\bibitem[Kahn et al.(1979)]{heao1}Kahn, S.M., Linsky, J.L., Mason, K.O., et al. 1979, ApJ, 234, L111
\bibitem[Kashyap et al.(2002)]{flares1}Kashyap, V.L., Drake, J.J., G\"udel, M., Audard, M. 2002, ApJ, 580, 1118
\bibitem[Katsova et al.(2002)]{kat02}Katsova, M.M., Livshits, M.A., Schmitt, J.H.M.M. 2002, ASP Conf. Proc., 177, 515
\bibitem[Kenyon \& Hartmann(1995)]{jcorr}Kenyon, S.J., Hartmann, L. 1995,ApJS, 101, 117
\bibitem[Kirsch et al.(2004)]{kirsch}Kirsch, M.G.F., Altieri, B., Chen, B., et al. 2004, astro-ph/0407257
\bibitem[Laming et al.(1995)]{laming}Laming, J.M., Drake, J.J., Widing, K.G. 1995, ApJ, 443, 416
\bibitem[Maggio et al.(2004)]{maggio}Maggio. A., Drake, J.J., Kashyap, V, et al. 2004, ApJ, 613, 548
\bibitem[Naftilan et al.(1992)]{naf92}Naftilan, S.A., Sandmann, W.S., Pettersen, B.R. 1992, PASP, 104, 1045
\bibitem[Ness et al.(2002)]{helike}Ness, J.-U., Schmitt, J.H.M.M., Burwitz, V., et al. 2002, A\&A, 394, 911
\bibitem[Ness et al.(2004)]{helike1}Ness, J.-U., G\"udel, M., Schmitt, J.H.M.M., et al. 2004, A\&A, 427, 667
\bibitem[Parker(1988)]{parker}Parker, E.N. 1988, ApJ, 330, 474
\bibitem[Porquet et al.(2001)]{porquet}Proquet, D., Mewe, R., Dubau, J., et al. 2001, A\&A, 376, 1113
\bibitem[Raassen et al.(2003)]{atmic}Raassen, A.J.J., Mewe, R., Audard, M. and G\"udel, M. 2003, A\&A, 411, 509
\bibitem[Robrade et al.(2004)]{eqpeg}Robrade, J., Ness, J.-U., Schmitt, J.H.M.M. 2004, A\&A, 413, 317
\bibitem[Rosner et al.(1978)]{rtv}Rosner, R., Tucker, W.H., Vaiana, G.S, 1978, ApJ, 220, 643
\bibitem[Sanz-Forcada et al.(2003)]{ab1}Sanz-Forcada, J., Maggio, A., Micela, G. 2003, A\&A, 408, 1087
\bibitem[Smith et al.(2001)]{apec}Smith, R.K., Brickhouse, N.S., Liedahl, D.A., Raymond, J.C. 2001, ASP Conf. Series, 247, 161
\bibitem[Schmitt \& Rosso(1988)]{schmitt88}Schmitt, J.H.M.M., Rosso, C. 1988, A\&A, 191, 99
\bibitem[Schmitt et al.(1995)]{schmitt95}Schmitt, J.H.M.M., Fleming, T.A., \& Giampapa, M.S. 1995, ApJ, 450, 392
\bibitem[Schmitt \& Liefke(2004)]{schmitt04}Schmitt, J.H.M.M., Liefke, C. 2004, A\&A, 417, 651
\bibitem[Sciortino et al.(1999)]{sax}Sciortino, S., Maggio, M., Favata, F., Orlando, S. 1999, A\&A, 342, 502
\bibitem[Stauffer \& Hartmann(1986)]{stau86}Stauffer, J.R., Hartmann, L.W. 1986, ApJS, 61, 531
\bibitem[Vaiana et al.(1981)]{vai81}Vaiana, G.S., Cassinelli, J.P., Fabbiano, R., et al. 1981, ApJ, 244, 163
\bibitem[Zboril et al.(1998)]{rotabu}Zboril, M., Byrne, P.B., Rolleston, W.R.J. 1998, MNRAS, 299, 753
\bibitem[Zuckerman et al.(2001)]{zuck01}Zuckerman, B., Song, I., Bessell, M.S., Webb, R.A., ApJ 562, L87
\end{thebibliography}
\end{document}